\documentclass[aps,prd,superscriptaddress,showpacs,preprint,amsmath,amssymb]{revtex4}
\usepackage{graphicx, bm}
\usepackage[usenames]{color}

\usepackage{subcaption}
\begin{document}

\draft
\title{Study on the anomalous quartic $W^+W^-\gamma\gamma$ couplings of electroweak bosons in $e^-p$ collisions at the LHeC and the FCC-he}

\author{ A. Guti\'errez-Rodr\'{\i}guez\footnote{alexgu@fisica.uaz.edu.mx}}
\affiliation{\small Facultad de F\'{\i}sica, Universidad Aut\'onoma de Zacatecas\\
         Apartado Postal C-580, 98060 Zacatecas, M\'exico.\\}


\author{ M. A. Hern\'andez-Ru\'{\i}z\footnote{mahernan@uaz.edu.mx}}
\affiliation{\small Unidad Acad\'emica de Ciencias Qu\'{\i}micas, Universidad Aut\'onoma de Zacatecas\\
         Apartado Postal C-585, 98060 Zacatecas, M\'exico.\\}

\author{E. Gurkanli\footnote{egurkanli@sinop.edu.tr}}
\affiliation{\small Department of Physics, Sinop University, Turkey.\\}

\author{V. Ari\footnote{vari@science.ankara.edu.tr}}
\affiliation{\small Department of Physics, Ankara University, Turkey.\\}

\author{M. K\"{o}ksal\footnote{mkoksal@cumhuriyet.edu.tr}}
\affiliation{\small Deparment of Optical Engineering, Sivas Cumhuriyet University, 58140, Sivas, Turkey.}

\date{\today}

\begin{abstract}

In this paper, a study is carried out on the $e^-p \to e^-\gamma^* p \to p W^-\gamma \nu_e$ production to probe quartic $W^+W^-\gamma\gamma$
couplings at the Large Hadron electron Collider (LHeC) with $\sqrt{s}$= 1.30, 1.98 TeV and at the Future Circular Collider-hadron electron (FCC-he)
with  $\sqrt{s}$= 3.46, 5.29 TeV. Production cross-sections are determined for both at leptonic and hadronic decay channels of the $W$-boson.
With the data from future $e^-p$ colliders, it is possible to obtain sensitivity measures at $95\%$ C.L. on the anomalous $f_ {M,i}/\Lambda^4$
and $ f_ {T,j}/\Lambda^4$ couplings which are competitive with the limits obtained by the LHC, as well as with others limits reported in the
literature. The production mode $e^-p \to e^-\gamma^* p \to p W^-\gamma \nu_e $ in $e^-p$ collisions offers a window for study the quartic $W^+W^-\gamma\gamma$ electroweak bosons couplings at the LHeC and the FCC-he, which provides a much cleaner collision environment than the LHC.

\end{abstract}

\pacs{12.60.-i, 14.70.Fm, 4.70.Bh  \\
Keywords: Models beyond the standard model, W bosons, Quartic gauge boson couplings.}

\vspace{5mm}

\maketitle

\section{Introduction}

A property of the weak interaction is that its gauge bosons $W^\pm$ and $Z$ can couple to each other in certain combinations
and also to $\gamma$. The gauge bosons $W^\pm$, $Z$, and $\gamma$ through mixing with each other represent some of the
Standard Model (SM) \cite{SM1,SM2,SM3} particles most strongly coupled to Electroweak Symmetry Breaking (EWSB). Due to
the non-Abelian nature of the SM electroweak theory, gauge bosons interact with each other and the SM predicts the existence
of the Triple Gauge Couplings (TGC) and the Quartic Gauge Couplings (QGC). While the anomalous TGC (aTGC) and the anomalous
QGC (aQGC) are deviations from the SM. Therefore, it is important to measure both aTGC and aQGC to further test the SM
or have indications of new physics Beyond the Standard Model (BSM). Study of anomalous $WW\gamma\gamma$ couplings sensitivity
is the main topic in this article. For this purpose, we use the effective Lagrangian formalism which has been utilized extensively
for parameterizing new physics BSM in many processes of particle physics. This technique provides a model-independent parameterization
of any new physics characterized by higher-order operators.

Studies for the aQGC $WW\gamma\gamma$ have been theoretically carried out at lepton-lepton colliders with the processes
$e^+e^- \to V V V$   \cite{Gutierrez,Belanger,Stirling,Leil,Dervan,Chong,Koksal,Chen}, $e^+e^- \to V V F F$
\cite{Stirling1,Senol-AHEP2017}, $e\gamma \to V V F$ \cite{Atag,Eboli}, $\gamma\gamma \to V V V$ \cite{Eboli1,Sahin},
$\gamma\gamma \to V V$ \cite{Koksal1}, $e^+e^- \to e^+ \gamma^* e^- \to V V F F$  \cite{Koksal2} and at hadron-hadron
colliders with the processes $pp \to V V V$ \cite{Yang,Eboli2,Bell,Ahmadov,Schonherr,Wen,Ye}, $pp \to V V F F$
\cite{Eboli4,Perez,Y-2002.04914}, $pp \to p\gamma^* p \to p V V F$ \cite{Y-2002.04914} and $pp \to p\gamma^*\gamma^* p \to pV V p$ \cite{Sahin1,Baldenegro,Fichet,Pierzchala,balde}, and $pp \to p\gamma^*\gamma^* p \to pV V V p$ \cite{tizchangg}, at lepton-hadron
colliders with the process $ep \to VVFF$ where $V=W^\pm, Z, \gamma$ and $F=e, j, \nu$. Searches for processes containing the aQGC have
been performed through the process $e^+e^- \to WW\gamma$ by the L3, DELPHI and OPAL Collaborations at the Large Electron Positron (LEP)
collider \cite{L3-Achard,DELPHI-Abdallah, OPAL-Abbiendi,OPAL-GAbbiendi}, the process $p\bar p \to pW^+W^-\bar p \to pe^+\nu e^-\bar\nu \bar p$
by the D0 Collaboration at the Tevatron \cite{D0-Abazov}, the processes $pp \to p\gamma^*\gamma^* p \to pWWp$ and $pp \to W\gamma jj$
by the CMS Collaboration \cite{CMS-Khachatryan,CMS-VKhachatryan} and the process $pp \to pW^+W^-p \to pe^\pm \nu\mu^\mp\nu p$ by the
ATLAS Collaboration at the LHC \cite{ATLAS-Aaboud-PRD}. In the post-LHC era the present and future colliders
contemplate in their physics programs the study of the aQGC: the High-Luminosity Large Hadron Collider (HL-LHC), the High-Energy
Large Hadron Collider (HE-LHC) \cite{HL-HE-LHC}, the Large Hadron electron Collider (LHeC) \cite{Fernandez,Fernandez1,Fernandez2,
LHeC-Bruning,LHeC-Web,LHeC-FCChe-2020,FCChe,Link-FCC-he-CERN}, the Future Circular Collider-hadron electron (FCC-he) \cite{FCChe,Link-FCC-he-CERN},
the International Linear Collider (ILC) \cite{ILC-Brau}, the Compact Linear Collider (CLIC) \cite{CLIC-Burrows}, the Circular
Electron Positron Collider (CEPC) \cite{CEPC-Ahmad} and the Future Circular Collider $e^+e^-$ (FCC-ee) \cite{TLEP-Bicer}.

The LHC may not provide high precision measurements due to strong interactions of $pp$ collisions. An $ep$ collider may be a very good
option to complement the LHC physics program. Since $ep$ colliders have high luminosity and high energy, the effects of new physics BSM
may appear by probing the interaction of $W$-boson with the photon which requires measuring $WW\gamma \gamma$ couplings precisely. The
LHeC and the FCC-he are planned to produce $ep$ collisions at energies from 1.30 TeV to 5.29 TeV \cite{Fernandez,Fernandez1,Fernandez2,LHeC-Bruning,LHeC-Web,LHeC-FCChe-2020}. The LHeC is a suggested deep inelastic electron-nucleon
scattering machine which has been planned to collide electrons with energy from 60 GeV to possibly 140 GeV, with protons with an energy
of 7 TeV. In addition, the FCC-he is designed to collide electrons with energy from 60 GeV to 140 GeV, with protons with an energy of 50 TeV.

In this paper, we present our results in a model-independent way for the total cross-section of the process $e^-p \to e^-\gamma^* p
\to p W^-\gamma \nu_e$ at the $e^-\gamma^*$ mode, as well as limits on the aQGC $WW\gamma\gamma$ at the LHeC with $\sqrt{s}$=1.30,
1.98 TeV and ${\cal L}=10, 100$ ${\rm fb^{-1}}$ and at the FCC-he with $\sqrt{s}$=3.46, 5.29 TeV and ${\cal L}=100, 1000$
${\rm fb^{-1}}$. For our study, we use an effective Lagrangian approach which provides a generic platform for introducing the effect
of new physics BSM by adding additional terms in the Lagrangian of the SM. Specifically, we consider a scenario where the electroweak
theory is realized linearly and the lowest-order aQGC are given by dimension-eight operators, and with a focus on the so-called genuine
aQGC operators, that is, operators that generate the aQGC but do not have any aTGC associated with them \cite{Eboli3}.

The paper is organized as follows: In Section II, we give the general expressions for the effective Lagrangian. In Section III,
we give a motivation on photon-induced process at future $ep$ colliders. In Section IV, we evaluate the total cross-section
of the reaction $e^-p \to e^-\gamma^* p \to p W^-\gamma \nu_e$. In Section V, we derive the $95\%$ C.L. allowed sensitivity
measures on the anomalous $f_ {M,i}/\Lambda^4$ and $ f_ {T,j}/\Lambda^4$ couplings at the LHeC and the FCC-he. In Section VI,
we summarize our conclusions.

\section{Dimension-8 operators set relevant for the process $e^-p \to e^-\gamma^* p \to p W^-\gamma \nu_e$}

A suitable and relatively modern approach to observe the effects of new physics BSM in a model-independent formalism
is to use an effective Lagrangian description of the SM.

Starting from our present theoretical, phenomenological and experimental understanding, treating the SM in an effective
Lagrangian approach is a well-motivated starting point since we have no present evidence of BSM physics.
In practice, this means defining a scale, $\Lambda$, of new physics higher than the energy scale being probed in the experiment
and using the fields of the SM to write higher dimension operators in addition to dimension-4 operators of the SM.
Following the context of Refs. \cite{LHeC-FCC-he-WWgg-Ari1,LHeC-FCC-he-WWgg-Ari2,LHeC-FCC-he-WWgg-Gurkanli,Eboli3,Degrande},
effective field theory in which the SM is extended by higher-dimensional operators composing by all possible combinations
of the SM fields is given by:

\begin{equation}
{\cal L}_{EFT}={\cal L}_{SM}+\sum_{i}\frac{c_i^{(6)}}{\Lambda^2}{\cal O}_i^{(6)}
+\sum_{j}\frac{c_j^{(8)}}{\Lambda^4}{\cal O}_j^{(8)}+...,
\end{equation}

\noindent Here, only even-dimension operators can contribute if we require lepton and baryon number conservation. For this reason,
the leading effective operators which give contribution to vertices including multi-bosons are expected from dimension-6
operators. Gauge boson operators have been described by either linear or non-linear effective Lagrangians. In the nonlinear
approach, the SM gauge symmetry is conserved and is realized by using the chiral Lagrangian parameterization \cite{Stirling,
PLB288-1992}. The aTGC and aQGC in this approach appear as dimension-6 operators. However, the SM gauge symmetry in the linear
approach is broken by means of Higgs scalar doublet \cite{Stirling,Eboli}. Generally, dimension-6 operators used to examine the
QGC provide great convenience for comparing LEP results \cite{Belanger}. Therefore, C and P conserving non-linear effective
Lagrangian for $WW\gamma\gamma$ aQGC that define dimension-6 operators is described as follows:

\begin{eqnarray}
{\cal L}_0&=& -\frac{e^2}{16\pi \Lambda^2} a_0 F_{\mu\nu}F^{\mu\nu}{\bf W}^\alpha {\bf W}_\alpha,   \\
{\cal L}_c&=& -\frac{e^2}{16\pi \Lambda^2} a_c F_{\mu\alpha}F^{\mu\beta}{\bf W}^\alpha {\bf W}_\beta,
\end{eqnarray}

\noindent where $F_{\mu\nu}$ and ${\bf W}^\alpha$ are defined in the usual way as in SM.

In our study in order to separate the effects of the aQGC, we shall consider effective operators that lead to the aQGC
without an aTGC associated to them. The lowest dimension operators that leads to quartic interactions are of dimension-8.
Therefore, genuine quartic vertices are of dimension-8. These operators are three classes of genuine aQGC operators and
they are given in Table I. This type of genuine aQGC operators in which we are interested do not have an aTGC counterpart.
In the set of genuine aQGC operators given in Table I, $\Phi$ stands for the Higgs doublet, and the covariant derivatives
of the Higgs field is given by $D_\mu\Phi=(\partial_\mu + igW^j_\mu \frac{\sigma^j}{2} + \frac{i}{2}g'B_\mu )\Phi$, and
$\sigma^j (j=1,2,3)$ represent the Pauli matrices, while $W^{\mu\nu}$ and $B^{\mu\nu}$ are the gauge field strength tensors
for $SU(2)_L$ and $U(1)_Y$, respectively.

It is worth mentioning that the LEP2 constraints on the $WW\gamma\gamma$ vertex \cite{Data2020} described in terms of the anomalous
$a_0/\Lambda^2$ and $a_c/\Lambda^2$ couplings can be translated into bounds on $f^2_{M,i}$ with $i=0-7$. The following expressions
show the relations between the $\frac{f_{M, i}}{\Lambda^4}$ couplings for the $WW\gamma\gamma$ vertex and the $\frac{a_0}{\Lambda^2}$
and $\frac{a_c}{\Lambda^2}$ couplings \cite{Belanger,Eboli3,Baak}:

\begin{eqnarray}
\frac{f_{M, 0}}{\Lambda^2}&=&\frac{a_0}{\Lambda^2}\frac{1}{g^2v^2},  \\
\frac{f_{M, 1}}{\Lambda^2}&=&-\frac{a_c}{\Lambda^2}\frac{1}{g^2v^2},  \\
\frac{f_{M, 0}}{\Lambda^2}&=&\frac{f_{M, 2}}{2}= \frac{f_{M, 6}}{2},    \\
\frac{f_{M, 1}}{\Lambda^2}&=&\frac{f_{M, 3}}{2}= -\frac{f_{M, 5}}{2}=\frac{f_{M, 7}}{2}.
\end{eqnarray}

To complement this section, in Table II, all the aQGC altered with dimension-8 operators are presented.

\begin{table}
\caption{Set of genuine aQGC operators of dimension-8 for $WW\gamma\gamma$ vertex \cite{Eboli3}.}
\begin{center}
\begin{tabular}{|c|c|}
\hline
\hline
Operator name  & Operator                         \\
\hline
\hline
\multicolumn{2}{|c|}{S-type operators}\\
\hline
$O_{S, 0}$     & $[(D_\mu\Phi)^\dagger (D_\nu\Phi)]\times [(D^\mu\Phi)^\dagger (D^\nu\Phi)]$     \\
\hline
$O_{S, 1}$     &  $[(D_\mu\Phi)^\dagger (D^\mu\Phi)]\times [(D_\nu\Phi)^\dagger (D^\nu\Phi)]$    \\
\hline
\hline
\multicolumn{2}{|c|}{M-type operators}    \\
\hline
$O_{M, 0}$   & $Tr[W_{\mu\nu} W^{\mu\nu}]\times [(D_\beta\Phi)^\dagger (D^\beta\Phi)]$    \\
\hline
$O_{M, 1}$   & $Tr[W_{\mu\nu} W^{\nu\beta}]\times [(D_\beta\Phi)^\dagger (D^\mu\Phi)]$  \\
\hline
$O_{M, 2}$   & $[B_{\mu\nu} B^{\mu\nu}]\times [(D_\beta\Phi)^\dagger (D^\beta\Phi)]$  \\
\hline
$O_{M, 3}$   & $[B_{\mu\nu} B^{\nu\beta}]\times [(D_\beta\Phi)^\dagger (D^\mu\Phi)]$  \\
\hline
$O_{M, 4}$   & $[(D_\mu\Phi)^\dagger W_{\beta\nu} (D^\mu\Phi)]\times B^{\beta\nu}$  \\
\hline
$O_{M, 5}$   & $[(D_\mu\Phi)^\dagger W_{\beta\nu} (D^\nu\Phi)]\times B^{\beta\mu}$  \\
\hline
$O_{M, 6}$   & $[(D_\mu\Phi)^\dagger W_{\beta\nu} W^{\beta\nu} (D^\mu\Phi)]$  \\
\hline
$O_{M, 7}$   & $[(D_\mu\Phi)^\dagger W_{\beta\nu} W^{\beta\mu} (D^\nu\Phi)]$  \\
\hline
\hline
\multicolumn{2}{|c|}{T-type operators}\\
\hline
$O_{T, 0}$   &  $Tr[W_{\mu\nu} W^{\mu\nu}]\times Tr[W_{\alpha\beta}W^{\alpha\beta}]$  \\
\hline
$O_{T, 1}$   &  $Tr[W_{\alpha\nu} W^{\mu\beta}]\times Tr[W_{\mu\beta}W^{\alpha\nu}]$  \\
\hline
$O_{T, 2}$   &  $Tr[W_{\alpha\mu} W^{\mu\beta}]\times Tr[W_{\beta\nu}W^{\nu\alpha}]$  \\
\hline
$O_{T, 5}$   &  $Tr[W_{\mu\nu} W^{\mu\nu}]\times B_{\alpha\beta}B^{\alpha\beta}$  \\
\hline
$O_{T, 6}$   &  $Tr[W_{\alpha\nu} W^{\mu\beta}]\times B_{\mu\beta}B^{\alpha\nu}$  \\
\hline
$O_{T, 7}$   &  $Tr[W_{\alpha\mu} W^{\mu\beta}]\times B_{\beta\nu}B^{\nu\alpha}$  \\
\hline
$O_{T, 8}$   &  $B_{\mu\nu} B^{\mu\nu}B_{\alpha\beta}B^{\alpha\beta}$  \\
\hline
$O_{T, 9}$   &  $B_{\alpha\mu} B^{\mu\beta}B_{\beta\nu}B^{\nu\alpha}$   \\
\hline
\end{tabular}
\end{center}
\end{table}

\begin{table}
\caption{The aQGC altered with dimension-8 operators are shown with X.}
\begin{center}
\begin{tabular}{|l|c|c|c|c|c|c|c|c|c|}
\hline\hline
& $WWWW$ & $WWZZ$ & $ZZZZ$ & $WW\gamma Z$ & $WW\gamma \gamma$ & $ZZZ\gamma$ & $ZZ\gamma \gamma$ & $Z \gamma\gamma\gamma$ & $\gamma\gamma\gamma\gamma$ \\
\hline
\cline{1-10}
$O_{S0}$, $O_{S1}$                     & X & X & X &   &   &   &   &   &    \\
$O_{M0}$, $O_{M1}$, $O_{M6}$, $O_{M7}$ & X & X & X & X & X & X & X &   &    \\
$O_{M2}$, $O_{M3}$, $O_{M4}$, $O_{M5}$ &   & X & X & X & X & X & X &   &    \\
$O_{T0}$, $O_{T1}$, $O_{T2}$           & X & X & X & X & X & X & X & X & X  \\
$O_{T5}$, $O_{T6}$, $O_{T7}$           &   & X & X & X & X & X & X & X & X  \\
$O_{T8}$, $O_{T9}$                     &   &   & X &   &   & X & X & X & X  \\
\hline
\hline
\end{tabular}
\end{center}
\end{table}

\begin{table}
\caption{Definition of the fiducial regions of the fully leptonic and hadronic $W^-\gamma \nu_e$ analyses.}
\begin{center}
\begin{tabular}{|ccc|}
\hline
\multicolumn{3}{|c|}{Fiducial Requirements}\\
\hline\hline
\multicolumn{3}{|c|}{Selected cuts of $f_{M,i}$ for the hadronic decay of the $W$-boson}\\
\hline
$p^j_T > 30$ \mbox{GeV},    &   $p^\gamma_T > 150$ \mbox{GeV},              &  $p^\nu_T > 20$ \mbox{GeV}      \\
        \qquad    \qquad       \qquad       100 \mbox{GeV} $>  M_{jj} > 60$ \mbox{GeV}, &
$|\eta_j| < 5$,             &  $|\eta_\gamma| < 2.5$, \\ $\Delta R_{jj (min)} = 0.4$   &  &  \\   $\Delta R_{\gamma j (min)} = 0.4$  & &           \\

\hline\hline
\multicolumn{3}{|c|}{Selected cuts of $f_{T,i}$ for the hadronic decay of the $W$-boson}\\
\hline
$p^j_T > 30$ \mbox{GeV},    &   $p^\gamma_T > 170$ \mbox{GeV},              &   $p^\nu_T > 20$ \mbox{GeV}     \\
        \qquad    \qquad       \qquad       100 \mbox{GeV} $>  M_{jj} > 60$ \mbox{GeV}, &

$|\eta_j| < 5$,             &  $|\eta_\gamma| < 2.5$,                \\ $\Delta R_{jj (min)} = 0.4$   &  &  \\   $\Delta R_{\gamma j (min)} = 0.4$  & &           \\

\hline
\hline
\multicolumn{3}{|c|}{Selected cuts of $f_{M,i}$ for the leptonic decay of the $W$-boson}\\
\hline
$p^l_T > 20$ \mbox{GeV},    &   $p^\gamma_T > 150$ \mbox{GeV},              & \quad $p^\nu_T > 20$ \mbox{GeV},       \\
$|\eta_l| < 2.5$,           &  $|\eta_\gamma| < 2.5$,  &             \\ $\Delta R_{ l l (min)} = 0.4$  &&  \\ $\Delta R_{\gamma l (min)} = 0.4$    &&               \\
\hline
\hline
\multicolumn{3}{|c|}{Selected cuts of $f_{T,i}$ for the leptonic decay of the $W$-boson}\\
\hline
$p^l_T > 20$ \mbox{GeV},    &   $p^\gamma_T > 170$ \mbox{GeV},              & \quad $p^\nu_T > 20$ \mbox{GeV},       \\
$|\eta_l| < 2.5$,           &  $|\eta_\gamma| < 2.5$,              &   \\ $\Delta R_{l l (min)} = 0.4$  && \\ $\Delta R_{\gamma l (min)} = 0.4$    &&             \\
\hline
\end{tabular}
\end{center}
\end{table}

\section{Photoproduction at the LHeC and the FCC-he}

Photon interactions have been extensively studied at HERA \cite{HERA}, LEP \cite{LEP}, Tevatron \cite{Abazov} and LHC \cite{Akiba}.
In a similar manner, a significant fraction of lepton-hadron collisions at the LHeC and the FCC-he will involve quasi-real photon
interactions. The LHeC and the FCC-he can to some extend be considered as a high energy $e\gamma^*$, $\gamma p$, $\gamma^* p$
and $\gamma^*\gamma^*$ collisions. On this topic, the future lepton-hadron colliders offer excellent new opportunities for the study
of high energy particle collisions, thus significantly extending the physics capabilities of a lepton-hadron collider.
With these options, a large number of new and exciting measurements become accessible with $e\gamma^*$, $\gamma p$, $\gamma^* p$
and $\gamma^*\gamma^*$ collisions. Because the photons couple directly to all fundamental fields carrying the electromagnetic currents
as leptons, quarks, $W'$s, etc.. High energy $e\gamma^*$, $\gamma p$, $\gamma^* p$ and $\gamma^*\gamma^*$ collisions will provide
a comprehensive laboratory for exploring virtually every aspect of the SM and BSM physics. A review of the studies made on $e\gamma^*$,
$\gamma p$, $\gamma^* p$ and $\gamma^*\gamma^*$ collisions physics on future colliders is made in Refs.
\cite{Ginzburg,Ginzburg1,Ginzburg-NPB1983,Zafer,Acar-NIMPRA2017,Aksacal-NIMPRA2007,Gutierrez-PRD2015,Billur-PRD2017,Koksal-IJMPA2019,Billur-EPJP2019,
Gutierrez-1903.04135,Billur-1909.10299,Gutierrez-JPG2020,Koksal-1910.06747,LHeC-FCC-he-WWgg-Ari1,LHeC-FCC-he-WWgg-Ari2,LHeC-FCC-he-WWgg-Gurkanli}.

It is appropriate to mention that the studies of photon interactions at the LHC are possible due to experimental signatures of events
involving photon exchanges such as the presence of very forward scattered protons and of large rapidity gaps in forward directions.
However, to tag efficiently photon-induced processes and to keep backgrounds under control, some  processes require very forward proton
detectors \cite{Piotrzkowski-PRD2001}. The photon-induced processes have been measured in $p\bar p$ collisions at Tevatron-Fermilab using
the large rapidity gap signature. The exclusive two-photon production of lepton pairs and  the diffractive photoproduction of $J/\psi$
mesons were studied in Refs. \cite{Abulencia-PRL2007,Aaltonen-PRL2009,Aaltonen1-PRL2009}, respectively. In both cases, clear signals were
obtained with low backgrounds.

As we mentioned above, scenarios like the LHeC and the FCC-he offer an unique opportunity to build $ep$ collider, which can also be operated
in $\gamma p$ collisions \cite{Ginzburg,Ginzburg1,Ginzburg-NPB1983,Zafer,Acar-NIMPRA2017,Aksacal-NIMPRA2007}. These conversions are made by
converting the incoming electrons or protons into an intense beam of high energy photons. In addition, the $ep$ colliders also provide the
opportunity to examine $\gamma^*\gamma^*$, $\gamma^*e$ and $\gamma^* p$ modes with quasi-real photons through the Equivalent Photon Approximation
(EPA) \cite{Budnev,Baur1,Piotrzkowski-PRD2001}.

The phenomenological investigations at lepton-hadron colliders generally contain usual deep inelastic scattering reactions
where the colliding hadron dissociates into partons. These reactions have been extensively studied in the literature, while the processes
elastic and semi-elastic, such as $e\gamma^*$, $\gamma^* \gamma^*$ and $\gamma^* p$ have been much less studied. These processes have simpler final
states with respect to lepton-hadron processes. In this case, these processes compensate for the advantages of lepton-hadron processes
such as high luminosity and high center-of-mass energy. In addition, $e\gamma^*$ have effective luminosity and much higher energy
compared to the process $\gamma^* \gamma^*$ collision. This may be significant because of the high energy dependencies of the cross-section
containing the new physics parameters. For all the aforementioned, it is expected that the $\gamma^* e$ collisions to have a high sensitivity
to the aQGC $WW\gamma\gamma$.

Regarding $e\gamma^*$ collisions, these can be discerned from usual deep inelastic scattering collisions by means of two experimental
signatures \cite{Rouby}. The first signature is the forward large rapidity gap \cite{Aaltonen-PRL2009,Aaltonen1-PRL2009,Chatrchyan-JHEP2012,Chatrchyan1-JHEP2012}. Quasi-real photons have a low virtuality and scatter
with small angles from the beam pipe. As the transverse momentum carried by a quasi-real photon is small, photon-emitting protons should
also be scattered with small angles and exit the central detector without being detected. This causes a decreased energy deposit
in the corresponding forward region. As a result, one of the forward regions of the central detector has a significant lack of
energy. This defines the forward large-rapidity gap, and usual $ep$ deep inelastic collisions can be rejected by applying a selection
cut on this quantity. The second experimental signature is provided by the forward detectors \cite{Fernandez,Buniatyan,Li-PRD100}
which are capable to detect particles with a large pseudorapidity. When a photon emitted by a proton is scattered with a large
pseudorapidity, it exceeds the pseudorapidity coverage of the central detectors. In these processes, the proton can be detected
by the forward detectors provides a distinctive signal for $e\gamma^*$ collisions. In this regard, the LHeC Collaboration has a
program of forward physics with extra detectors located in a region between a few tens up to several hundreds of metres from the
interaction point \cite{Fernandez}.

\section{The total cross-section for one exchanged quasi-real photon}

$\gamma^*$ photons emitted from proton beams collide with the incoming electron, and $e\gamma^*$ collisions are generated.
The process $e^-\gamma^* \to W^-\gamma \nu_e$ participates as a subprocess in the process $e^-p \to e^-\gamma^* p \to p W^-\gamma \nu_e$.
In addition, the diagram of the process $e^-p \to e^-\gamma^* p \to p W^-\gamma \nu_e$ is given in Fig. 1. The Feynman diagrams for the
subprocess $e^-\gamma^* \to W^-\gamma \nu_e$ are shown in Fig. 2. Therefore, we find the total cross-section of the main process
$e^-p \to e^-\gamma^* p \to p W^-\gamma \nu_e$ by integrating the cross-section for the subprocess  $e^-\gamma^* \to W^-\gamma \nu_e$.
The total cross-section of this process can be written as:

\begin{equation}
\sigma(e^-p \to p W^-\gamma \nu_e) = \int f_{\gamma^*}(x){\hat\sigma}(e^-\gamma^* \to W^-\gamma \nu_e) dx.
\end{equation}

\noindent \newline Here, the spectrum of EPA photons $f_{\gamma^*}(x)$ is defined as follows \cite{Budnev,Belyaev}:

\begin{eqnarray}
f_{\gamma^*}(x)=\frac{\alpha}{\pi E_{p}}\{[1-x]\left[\varphi(\frac{Q_{max}^{2}}{Q_{0}^{2}})-\varphi(\frac{Q_{min}^{2}}{Q_{0}^{2}})\right]\},
\end{eqnarray}

\noindent \newline with $x=E_{\gamma}/E_{p}$, $Q^2_{max}=2\hspace{0.8mm}GeV^2$ is the maximum virtuality of the photon and $Q^2_{min}$ is:

\begin{eqnarray}
Q_{min}^{2}=\frac{m_{p}^{2}x^{2}}{1-x}.
\end{eqnarray}

\noindent \newline In addition, the explicit form of function $\varphi$ contained in Eq. (9) is:

\begin{eqnarray}
\varphi(\theta)=&&(1+ay)\left[-\textit{In}(1+\frac{1}{\theta})+\sum_{k=1}^{3}\frac{1}{k(1+\theta)^{k}}\right]
+\frac{y(1-b)}{4\theta(1+\theta)^{3}} \nonumber \\
&& +c(1+\frac{y}{4})\left[\textit{In}\left(\frac{1-b+\theta}{1+\theta}\right)+\sum_{k=1}^{3}\frac{b^{k}}{k(1+\theta)^{k}}\right],
\end{eqnarray}

\noindent \newline where explicitly $a$, $b$, $c$ and $y$ are given by:

\begin{eqnarray}
y=\frac{x^{2}}{(1-x)},
\end{eqnarray}

\begin{eqnarray}
a=\frac{1+\mu_{p}^{2}}{4}+\frac{4m_{p}^{2}}{Q_{0}^{2}}\approx 7.16,
\end{eqnarray}

\begin{eqnarray}
b=1-\frac{4m_{p}^{2}}{Q_{0}^{2}}\approx -3.96,
\end{eqnarray}

\begin{eqnarray}
c=\frac{\mu_{p}^{2}-1}{b^{4}}\approx 0.028.
\end{eqnarray}

In our calculations, we analyze signals and backgrounds of the process $e^-p \to e^-\gamma^* p \to p W^-\gamma \nu_e$ through
the expression given by Eq. (8) and using the MadGraph5\_aMC@NLO \cite{MadGraph} package in which the aQGC are implemented through
FeynRules \cite{AAlloul} package through dimension-8 effective Lagrangians related to the anomalous quartic $WW\gamma\gamma$ couplings.

In order to make our numerical computation more realistic, the kinematic study of the $W^-\gamma\nu_e$ production starts with the usual
detection and isolation cuts on the final state leptons and quarks. The SM process with final state should be accepted as a background
for the process $e^-p \to e^-\gamma^* p \to p W^-\gamma \nu_e$. Additionally, we have considered the following background processes $\nu_{e} \gamma l \nu_{l}$, 
$\nu_{e}\gamma jj$, $\nu_e \gamma \gamma jj$ and $\nu_e \gamma jjj$ for both leptonic and hadronic decay channel of the W-boson. We know that the high dimensional operators could affect
the $p^{\gamma}_T$ photon transverse momentum, especially at the region with a large $p^{\gamma}_T$ values, which can be very
useful to distinguish signal and background events (see Section V). By applying a cut in the missing energy, we reduce the background
to consider. Therefore, a set of kinematic cuts used for the analysis of signal and background processes are summarized in Table III.

In Tables IV and V, we present the values of $\sigma_{SM}$ and $\sigma_{Tot}$ of the process $e^-p \to e^-\gamma^* p \to p W^-\gamma \nu_e$ for the center-of-mass energies of 1.30, 1.98, 3.46 and 5.29 TeV. Here, $\sigma_{Tot}$ is given as follows,

\begin{eqnarray}
\sigma_{Tot}=\sigma_{SM}+\sigma_{int}+\sigma_{BSM}
\end{eqnarray}
where $\sigma_{SM}$ is the SM cross-section, $\sigma_{int}$ is the interference term between SM and the new physics contribution and $\sigma_{BSM}$ is the contribution due to BSM physics, respectively. Values of the total cross-section predictions for the process $e^-p \to e^-\gamma^* p \to p W^-\gamma \nu_e$ after
applying the cuts described in the text, show that the total cross-section increases with increasing of the center-of-mass energy,
as well as with the anomalous $f_ {M,i}/\Lambda^4$ and $ f_ {T,j}/\Lambda^4$ contribution. For instance, from Tables IV
and V, we obtain $\sigma(\sqrt{s}, \frac{f_{T,6}}{\Lambda^4}) = (3.18\times 10^7)\sigma_{SM}$ and
$\sigma(\sqrt{s},\frac{f_{T,5}}{\Lambda^4}) = (2.66\times 10^7)\sigma_{SM}$ for the leptonic and hadronic channels,
with $\sqrt{s}= 5.29$ TeV at the FCC-he. In general, the total cross-section predicted for the
$e^-p \to e^-\gamma^* p \to p W^-\gamma \nu_e$ signal and for the center-of-mass energies of $\sqrt{s}= 1.30, 1.98, 3.46, 5.29$ TeV
is increased by ${\cal O}(10^1-10^7)$ orders of magnitude with respect to the total cross-section of the SM, that is to say $\sigma(\sqrt{s}, \frac{f_{M,i}}{\Lambda^4})\approx {\cal O}(10^1-10^5)\sigma_{SM}$ and $\sigma(\sqrt{s}, \frac{f_{T,j}}{\Lambda^4})\approx {\cal O}(10^4-10^7)
\sigma_{SM}$, with $i= 1-5, 7$ and $j= 0-2, 5, 6, 7$.

For the aQGC $f_{M,0-5,7}/\Lambda^4$ and $f_{T,0-2,5,6,7}/\Lambda^4$ taking one at a time, we get the results for the total cross-section ($\sigma_{Tot}$)
as shown in Figs. 3-10 and Tables IV and V at the LHeC with $\sqrt{s}=1.30, 1.98$ TeV and the FCC-he with $\sqrt{s}=3.46, 5.29$ TeV,
respectively. The color lines in Figs. 3-10 show the deviations from the total SM value of the process $e^-p \to e^-\gamma^* p \to p W^-\gamma \nu_e$
as a function of $f_{M, i}/\Lambda^4$ and $f_{T, j}/\Lambda^4$. In these figures, we consider the leptonic and hadronic decays of the $W$-boson
in the final state of the process $e^-p \to e^-\gamma^* p \to p W^-\gamma \nu_e$, where $W \to \nu_l l$, $W \to jj'$ with $\nu_l=\nu_e, \nu_\mu$, $l=e^-, \mu$
and $j=u, c, \bar d, \bar s$, $j'=d, s, \bar u, \bar c$. Figs. 8 and 10 illustrate more clear effect of the dimension-8 operators on the total
cross-section of the process $e^-p \to e^-\gamma^* p \to p W^-\gamma \nu_e$ with the leptonic and hadronic decay of the $W$-boson, and for the
FCC-he with $ \sqrt{s}=5.29 $ TeV. The highest cross-section in value is obtained for $\sigma(\sqrt{s}, f_{T,5}/\Lambda^4) = 2.10 \times 10^{5}$ pb
followed by $\sigma(\sqrt{s}, f_{T,6}/\Lambda^4) = 1.05 \times 10^{5}$ pb and $\sigma(\sqrt{s}, f_{T,7}/\Lambda^4) = 2.54 \times 10^{4}$ pb for the
hadronic channel as shown in Table V as well as by Fig. 10. A direct comparison of the results shown in Tables IV
and V for the total cross-section of the process $e^-p \to e^-\gamma^* p \to p W^-\gamma \nu_e$ projected by the LHeC and the FCC-he
for both leptonic and hadronic channels of the $W$-boson, indicate a difference of one and up to two orders of magnitude of the FCC-he
with respect to the LHeC. Similar behavior is observed in Fig. 3-10.

To close this section, it is worth mentioning that our results show that a nonzero aQGC enhances the production cross-section at large
energies of the $e^-p$ system with respect to the SM prediction as can be seen in Figs. 3-10.

\section{Projections on the aQGC $f_{M,i}/\Lambda^4$ and $f_{T,j}/\Lambda^4$ at the LHeC and the FCC-he}

The presence of new physics characterized by the parameters $f_{M,i}/\Lambda^4$ and $f_{T,j}/\Lambda^4$ may be quantified by a simple
$\chi^2$ method that varies the parameters and is based on:

\begin{equation}
\chi^2(f_{M,i}/\Lambda^4, f_{T,j}/\Lambda^4)=\Biggl(\frac{\sigma_{SM}(\sqrt{s}) -\sigma_{Tot}(\sqrt{s}, f_{M,i}/\Lambda^4, f_{T,j}/\Lambda^4)}
{\sigma_{SM}(\sqrt{s})\delta_{st}}\Biggr)^2.
\end{equation}
where $\delta_{st}=\frac{1}{\sqrt{N_{SM}}}$ is the statistical error and $N_{SM}$ is the total number of events only coming from SM backgrounds:

\begin{equation}
N_{SM}={\cal L}\times \sigma_{SM}.
\end{equation}

In order to quantify the expected limits on $f_{M,i}/\Lambda^4$ and $f_{T,j}/\Lambda^4$, advantage has been taken in this analysis
of the fact that the aQGC enhance the total cross-section at high energies (see Tables IV and V). To get an idea of the LHeC and
FCC-he constraining power, in Tables VI and VII, we show the expected bounds on the aQGC $f_{M,i}/\Lambda^4$ and $f_{T,j}/\Lambda^4$ from
the $e^-p \to e^-\gamma^* p \to p W^-\gamma \nu_e$ production. We present in the rows of Tables VI and VII the expected LHeC and FCC-he
limits. In these tables, attainable sensitivity on $f_{M,0-5, 7}/\Lambda^4$ and $f_{T,0-2, 5-7}/\Lambda^4$ at the LHeC and the
FCC-he runs is already higher than the present direct limits stemming from LEP \cite{L3-Achard,DELPHI-Abdallah,OPAL-Abbiendi,OPAL-GAbbiendi}
and Tevatron \cite{D0-Abazov} and our limits are competitive with the limits reported by the LHC \cite{CMS-Khachatryan,CMS-VKhachatryan,ATLAS-Aaboud-PRD}.
At the FCC-he, these aQGC can be tested in the $e^-p \to e^-\gamma^* p \to p W^-\gamma \nu_e$ production mode with $\sqrt{s}= 5.29$ TeV and
${\cal L}=1000$ ${\rm fb^{-1}}$. Our limits stronger on the Wilson coefficients of the operators ${\cal O}_{T,j}$ are listed below for the FCC-he
with $\sqrt{s}=5.29$ TeV and ${\cal L}=1000$ ${\rm fb^{-1}}$ at $95\%$ C.L., for the hadronic decay channel
of the $W$-boson in the final state:

\begin{eqnarray}
\frac{f_{T5}}{\Lambda^{4}}&=& [-0.237; 0.270] \hspace{1mm} {\rm TeV^{-4}}, \\
\frac{f_{T6}}{\Lambda^{4}}&=& [-0.330; 0.420] \hspace{1mm} {\rm TeV^{-4}}, \\
\frac{f_{T7}}{\Lambda^{4}}&=& [-1.000; 0.550] \hspace{1mm} {\rm TeV^{-4}}.
\end{eqnarray}

For the operators ${\cal O}_{M,i}$, the constraints on the Wilson coefficients do not degrade substantially  and become  in a factor
${\cal O}(2-5)$ weaker than those obtained in Refs. \cite{CMS-2002.09902,Eduardo-2004.05174} through the $Z\gamma jj$ production.

To complement our results obtained on the anomalous $f_{M,i}/\Lambda^{4}$ and $f_{T,j}/\Lambda^{4}$ couplings which are summarized
in Tables VI and VII, we calculated the sensitivity on the aQGCs at the $95\%$ C. L. via the process $e^- p \to e^-\gamma^* p \to
p W^- \gamma \nu_e$ at the FCC-he with $\sqrt{s}$ = 3.46, 5.29 TeV with the combined data for the leptonic and hadronic decays
of $W$-boson given in Table III. Our results are illustrated in Table VIII. In this case the sensitivities on the aQGCs for the
combined data get better with respect to the leptonic and hadronic decays of $W$-boson given by Tables VI and VII.

In the post-LHC era, the FCC-he is one of the proposed colliders in the new energy frontier at the LHC and would provide
proton beam energies up to 50 TeV and electron beam energies from  60 GeV to 140 GeV. In this case the expected bounds
on the $f_{M,0-5, 7}/\Lambda^4$ and $f_{T,0-2, 5-7}/\Lambda^4$ Wilson coefficients can reach a sensitivity of approximately
one order of magnitude stronger than our present limits given in Tables VI-VIII.

In Ref. \cite{Eduardo-2004.05174}, a study phenomenological through the two-to-two scattering of electroweak gauge bosons is carried
out to determined the partial-wave unitarity constraints on the lowest-dimension effective operators which generate aQGC. Quantitatively,
its results are summarized in Tables I-IV of Ref. \cite{Eduardo-2004.05174}. Our results on the $f_{M,i}/\Lambda^4$ and $f_{T,j}/\Lambda^4$
Wilson coefficients show that our limits are competitive with the results reported in Ref. \cite{Eduardo-2004.05174}, and in some cases
our limits are stronger by one order of magnitude as shown in Tables VII and VIII of our manuscript, as well as in Tables II and III of
Ref. \cite{Eduardo-2004.05174}.

Our results for the anomalous $f_{M,i}/\Lambda^4$ and $f_{T,j}/\Lambda^4$ couplings are competitive with those reported in
Ref. \cite{CMS-2002.09902} through the $Z\gamma jj$ production at $\sqrt{s}= 13$ TeV and integrated luminosity of 35.9 $\rm fb^{-1}$
at the LHC. A direct comparison of the anomalous $f_{M,i}/\Lambda^4$ and $f_{T,j}/\Lambda^4$ couplings given in Ref. \cite{CMS-2002.09902}
with our results reported in Tables VI-VIII, shows that in some cases our bounds for $f_{M,i}/\Lambda^4$ and $f_{T,j}/\Lambda^4$
are more stringent than those reported in Table IV of Ref. \cite{CMS-2002.09902} for the CMS Collaboration at the LHC. In Ref. \cite{CMS-PLB2019},
a search at $95\%$ C.L. for the aQGC $f_{M,0, 1, 6, 7}/\Lambda^4$ and $f_{T,0, 1, 2}/\Lambda^4$ through electroweak production of $WW$,
$WZ$, and $ZZ$ boson pairs in association with two jets in proton-proton collisions at the LHC with $\sqrt{s}=13$ TeV is reported.
Our results are competitive with those reported by Ref. \cite{CMS-PLB2019}.

Other experimental results on the anomalous $f_{M,i}/\Lambda^4$ and $f_{T,j}/\Lambda^4$ couplings reported by the CMS and ATLAS
Collaborations are the followings. With $\sqrt{s} = 8$ TeV and integrated luminosity of 19.7 $\rm fb^{-1}$ the CMS experiment
searching for exclusive or quasi-exclusive $WW$ production via the signal topology $pp \to p\gamma^* \gamma^* p \to pW^+W^-p $ \cite{CMS-Khachatryan,CMS-VKhachatryan}. Their research are translated into limits on the aQGC $f_{M,0, 1, 2, 3}/\Lambda^4$.
In addition, the CMS experiment measure the electroweak-induced production of $W$ and two jets, where the $W$-boson decays
leptonically, and experimental limits on the aQGC $f_{M,0-7}/\Lambda^4$ and $f_{T,0-2, 5-7}/\Lambda^4$ are set at $95\%$ C.L.
\cite{CMS-Khachatryan}. In another investigation with $\sqrt{s} = 8$ TeV and ${\cal L}=20.2$ ${\rm fb^{-1}}$ of $pp$ collisions
the ATLAS experiment, was studied the production of $WV\gamma$ events in $e\nu\mu\nu\gamma$, $e\nu j j \gamma$ and $\mu\nu jj\gamma$
final states \cite{ATLAS-EPJC2017}. The results reported in these studies are weaker than those reported in our present article.

To be consistent and put in perspective the current limits with a small luminosity compared to the end run of the HL-LHC,
as well as with  the projections of limits at future electron-proton colliders with full luminosity, we present the expected
bounds on the anomalous quartic $WW\gamma\gamma$ couplings $f_{M,0-1}/\Lambda^4$ and $f_{T,0-1}/\Lambda^4$ at the HL-LHC with
$\sqrt{s}=14$ TeV and ${\cal L}=3000$ ${\rm fb}^{-1}$:

\begin{eqnarray}
\frac{f_{M0}}{\Lambda^{4}}&=& [-4.0; 4.0] \hspace{1mm} {\rm TeV^{-4}}, \\
\frac{f_{M1}}{\Lambda^{4}}&=& [-12.0;  12.0] \hspace{1mm} {\rm TeV^{-4}}, \\
\frac{f_{T0}}{\Lambda^{4}}&=& [-0.6; 0.6] \hspace{1mm} {\rm TeV^{-4}},   \\
\frac{f_{T1}}{\Lambda^{4}}&=& [-0.4; 0.4] \hspace{1mm} {\rm TeV^{-4}}.
\end{eqnarray}

\noindent These results show that even in the case of the HL-LHC, our limits on the aQGC given in Tables VII and VIII
are competitive and in some cases are better with respect to the limits given by Eqs. (22)-(25).

Phenomenological results on the aQGC $f_{M,i}/\Lambda^4$ and $f_{T,j}/\Lambda^4$ at the LHeC with $\sqrt{s}$ = 1.30, 1.98 TeV and the FCC-he
with $\sqrt{s}$ = 3.46, 5.29 TeV are presented in Refs. \cite{LHeC-FCC-he-WWgg-Ari1,LHeC-FCC-he-WWgg-Ari2,LHeC-FCC-he-WWgg-Gurkanli}.
In Ref. \cite{LHeC-FCC-he-WWgg-Gurkanli}, the $ep \to e^-\gamma^* p \to eW\gamma q'X \to e\nu_l lq'X$ channel gets sensitivity measures
of the order of $10^{-1}$ for some anomalous $f_{T,j}/\Lambda^4$ couplings. In another study, sensitivity measures on the aQGC of the order
of $10^1$ are reported by Refs. \cite{LHeC-FCC-he-WWgg-Ari1,LHeC-FCC-he-WWgg-Ari2} via the process $e^-p \to e^-\gamma^*\gamma^*p \to e^-W^+W^-p$
with the subprocess $\gamma^*\gamma^* \to W^+W^-$. In the case of our present article, in Tables VI-VIII, we summarize all of the sensitivity
measures on the anomalous $f_{M,i}/\Lambda^4$ and $f_{T,j}/\Lambda^4$ couplings obtained at $\sqrt{s}$ = 1.30, 1.98 TeV and $\sqrt{s}$ = 3.46, 5.29 TeV
with the production mode $e^-p \to e^-\gamma^* p \to p W^-\gamma \nu_e $. Our results on the aQGC $f_{M,0-5,7}/\Lambda^4$ and $f_{T,0-2,5-7}/\Lambda^4$
for the different energy stages above mentioned provided sensitivity measures of the order of $10^{-1}$, which is similar to those sensitivity measures
report by \cite{LHeC-FCC-he-WWgg-Ari1,LHeC-FCC-he-WWgg-Ari2,LHeC-FCC-he-WWgg-Gurkanli} at the LHeC and the FCC-he, with other channels.
For other reviews experimental and phenomenological, the reader can check Refs.
\cite{Liga-aQGC,CMS-Khachatryan,CMS-VKhachatryan,ATLAS-EPJC2017,LHeC-FCC-he-WWgg-Ari1,LHeC-FCC-he-WWgg-Ari2,
Yu-LHC2020.03326,Eduardo-2004.05174}.

\begin{table}
\caption{Total cross-sections of the process $e^- p \to e^-\gamma^* p \to p W^- \gamma \nu_e $ at the LHeC
with $\sqrt{s}=1.30, 1.98$ TeV and at the FCC-he with $\sqrt{s}=3.46, 5.29$ depending on 13 anomalous couplings
obtained by dimension-8 operators. Also, all anomalous couplings for the LHeC and the FCC-he are taken as equal
to $1\times 10^{-8}$ and $5\times 10^{-9}$ ${\rm GeV}^{-4}$, respectively. Here, we consider that only one of the
anomalous couplings deviates from the SM at any given time.}
\begin{center}
\begin{tabular}{|c|c|c|c|c|}
\hline\hline
\multicolumn{5}{|c|}{$\sigma(e^- p \to e^-\gamma^* p \to p W^- \gamma \nu_e )$ (pb)}\\
\hline \hline
     & \multicolumn{2}{|c|}{LHeC} & \multicolumn{2}{|c|}{FCC-he}\\
     & \multicolumn{2}{|c|}{Leptonic decay} & \multicolumn{2}{|c|}{Leptonic decay}\\
\hline
SM                    & $7.55 \times 10^{-6}$ & $4.04 \times 10^{-5}$  &  $3.93\times 10^{-5}$   & $1.63 \times 10^{-4}$   \\
\hline
Couplings             & $\sqrt{s}= 1.30$ TeV  & $\sqrt{s} = 1.98 $ TeV  &  $\sqrt{s} = 3.46 $ TeV & $\sqrt{s} = 5.29 $ TeV  \\
\hline
$f_{M0}/\Lambda^{4}$  & 2.99 $\times 10^{-3}$ & 9.35 $\times 10^{-2}$   & 1.09 $\times 10^{-2}$   &   7.77 $\times 10^{-1}$  \\
\hline
$f_{M1}/\Lambda^{4}$  & 2.04 $\times 10^{-3}$ & 5.70 $ \times 10^{-2}$  & 7.43 $\times 10^{-3}$   &   6.00 $\times 10^{-1}$  \\
\hline
$f_{M2}/\Lambda^{4}$  & 1.29 $\times 10^{-1}$ & 4.02                    & 4.73 $\times 10^{-1}$   &   3.25 $\times 10^{1}$ \\
\hline
$f_{M3}/\Lambda^{4}$  & 8.81 $\times 10^{-2}$ & 2.45                    & 3.23 $\times 10^{-1}$   &   2.58 $\times 10^{1}$ \\
\hline
$f_{M4}/\Lambda^{4}$  & 9.77 $\times 10^{-3}$ & 3.06 $\times 10^{-1}$   & 3.56 $\times 10^{-2}$   &   2.54 \\
\hline
$f_{M5}/\Lambda^{4}$  & 6.77 $\times 10^{-3}$ & 1.87 $\times 10^{-1}$   & 2.52 $\times 10^{-2}$   &   1.96 \\
\hline
$f_{M7}/\Lambda^{4}$  & 5.25 $\times 10^{-4}$ & 1.43 $\times 10^{-2}$   & 1.96 $\times 10^{-3}$   &   1.46 $\times 10^{-1}$ \\
\hline
$f_{T0}/\Lambda^{4}$  & 2.85 $\times 10^{-1}$ & 10.00                   & 3.03                    &   2.14 $\times 10^{2}$  \\
\hline
$f_{T1}/\Lambda^{4}$  & 7.15 $\times 10^{-1}$ & 2.34 $\times 10^{1}$    & 5.79                    &   4.82 $\times 10^{2}$  \\
\hline
$f_{T2}/\Lambda^{4}$  & 9.60 $\times 10^{-2}$  & 3.16                   & 8.18 $\times 10^{-1}$   &   7.75 $\times 10^{1}$  \\
\hline
$f_{T5}/\Lambda^{4}$  & 3.07                   & 1.08 $\times 10^{2}$   & 3.26 $\times 10^{1}$    &   2.26 $\times 10^{3}$ \\
\hline
$f_{T6}/\Lambda^{4}$  & 7.68                   & 2.52 $\times 10^{2}$   & 6.25 $\times 10^{1}$    &   5.19 $\times 10^{3}$\\
\hline
$f_{T7}/\Lambda^{4}$  & 1.04                   & 3.40 $\times 10^{1}$   & 8.79                   &   7.63 $\times 10^{2}$\\
\hline
\end{tabular}
\end{center}
\end{table}

\begin{table}
\caption{Total cross-sections of the process $e^- p \to e^-\gamma^* p \to p W^- \gamma \nu_e$ at the LHeC
with $\sqrt{s}=1.30, 1.98$ TeV and at the FCC-he with $\sqrt{s}=3.46, 5.29$ TeV depending on 13 anomalous
couplings obtained by dimension-8 operators. Also, all anomalous couplings for the LHeC and the FCC-he
are taken as equal to $1\times 10^{-8}$ and $5\times 10^{-9}$ ${\rm GeV}^{-4}$, respectively.
Here, we consider that only one of the anomalous couplings deviates from the SM at any given time.}
\begin{center}
\begin{tabular}{|c|c|c|c|c|}
\hline \hline
\multicolumn{5}{|c|}{$\sigma(e^- p \to e^-\gamma^* p \to p W^- \gamma \nu_e) $ (pb)}\\
\hline \hline
     & \multicolumn{2}{|c|}{LHeC} & \multicolumn{2}{|c|}{FCC-he}\\
     & \multicolumn{2}{|c|}{Hadronic decay} & \multicolumn{2}{|c|}{Hadronic decay}\\
\hline
SM                    & $1.98 \times 10^{-5}$ & $8.95 \times 10^{-5}$  &  $2.37\times 10^{-4}$   & $6.38 \times 10^{-4}$   \\
\hline
Couplings             & $\sqrt{s} = 1.30$ TeV & $\sqrt{s} = 1.98$ TeV  & $\sqrt{s}=3.46$ TeV   & $\sqrt{s}= 5.29$ TeV \\
\hline
$f_{M0}/\Lambda^{4}$  & 1.85 $\times 10^{-2}$ & 3.17 $\times 10^{-1}$ & 1.06                   & 8.33  \\
\hline
$f_{M1}/\Lambda^{4}$  & 5.66 $\times 10^{-3}$ & 1.03 $\times 10^{-1}$ & 9.16 $\times 10^{-2}$  & 8.04 $\times 10^{-1}$    \\
\hline
$f_{M2}/\Lambda^{4}$  & 7.94  $\times 10^{-1}$    & 1.36 $\times 10^{1}$  & 4.50 $\times 10^{1}$   & 3.58 $\times 10^{2}$ \\
\hline
$f_{M3}/\Lambda^{4}$  & 2.45 $\times 10^{-1}$ & 4.43                  & 3.90                   & 3.48 $\times 10^{1}$ \\
\hline
$f_{M4}/\Lambda^{4}$  & 6.05 $\times 10^{-2}$ & 1.04                  & 3.43                   & 2.90 $\times 10^{1}$   \\
\hline
$f_{M5}/\Lambda^{4}$  & 1.88 $\times 10^{-2}$ & 3.39 $\times 10^{-1}$ & 2.97 $\times 10^{-1}$  & 2.63  \\
\hline
$f_{M7}/\Lambda^{4}$  & 1.46 $\times 10^{-3}$ & 2.58 $\times 10^{-2}$  & 2.33 $\times 10^{-2}$  & 2.02 $\times 10^{-1}$\\
\hline
$f_{T0}/\Lambda^{4}$  & 1.30                  & 2.89 $\times 10^{1}$  & 1.48 $\times 10^{2}$   & 1.55 $\times 10^{3}$\\
\hline
$f_{T1}/\Lambda^{4}$  & 1.93                  & 4.19 $\times 10^{1}$  & 7.10 $\times 10^{1}$   & 7.47 $\times 10^{2}$\\
\hline
$f_{T2}/\Lambda^{4}$  & 2.79 $\times 10^{-1}$ & 6.03                  & 1.77 $\times 10^{1}$   & 1.87 $\times 10^{2}$\\
\hline
$f_{T5}/\Lambda^{4}$  & 1.40 $\times 10^{1}$  & 3.10 $\times 10^{2}$  & 1.59 $\times 10^{3}$   & 1.70 $\times 10^{4}$\\
\hline
$f_{T6}/\Lambda^{4}$  & 2.08 $\times 10^{1}$  & 4.51 $\times 10^{2}$  & 7.61 $\times 10^{2}$   & 8.07 $\times 10^{3}$\\
\hline
$f_{T7}/\Lambda^{4}$  & 3.01                  & 6.52 $\times 10^{1}$  & 1.94 $\times 10^{2}$   & 2.02 $\times 10^{3}$\\
\hline
\end{tabular}
\end{center}
\end{table}


\begin{table}
\caption{Sensitivity measures on the aQGC at the $95\%$ C. L. via the process $e^- p \to e^-\gamma^* p \to p W^- \gamma \nu_e$
at the LHeC with $\sqrt{s} = 1.30, 1.98$ TeV. Here, we consider that only one of the anomalous couplings deviates from the
SM at any given time.}
\begin{tabular}{|c|c|c|c|c|c|}
\hline \hline
\multicolumn{5}{|c|}{LHeC, \hspace{5mm} $\sqrt{s}$ = 1.30 TeV }\\
\hline
& \multicolumn{2}{|c|}{Leptonic channel} & \multicolumn{2}{|c|}{ Hadronic channel }\\
\hline
Couplings (TeV$^{-4}$) & 10 fb$^{-1}$                & 100 fb$^{-1}$                & 10 fb$^{-1}$                  & 100 fb$^{-1}$ \\
\hline
\cline{1-5}
$f_{M0}/\Lambda^{4}$  & [-1.45;1.25] $\times 10^{3}$ & [-0.86;0.66] $\times 10^{3}$ & [-6.75;6.92] $ \times 10^{2}$ & [-3.76;3.93] $ \times 10^{2}$ \\
\hline
$f_{M1}/\Lambda^{4}$  & [-2.02;2.10] $\times 10^{3}$ & [-0.88;0.94] $\times 10^{3}$ & [-1.24;1.26] $ \times 10^{3}$ & [-6.96;7.10] $ \times 10^{2}$ \\
\hline
$f_{M2}/\Lambda^{4}$  & [-2.02;2.10] $\times 10^{2}$ & [-1.12;1.20] $\times 10^{2}$ & [-1.02;1.06] $ \times 10^{2}$ & [-5.62;6.09] $ \times 10^{1}$ \\
\hline
$f_{M3}/\Lambda^{4}$  & [-2.39;2.55] $\times 10^{2}$ & [-1.31;1.47] $\times 10^{2}$ & [-1.86;1.93] $ \times 10^{2}$ & [-1.03;1.10] $ \times 10^{2}$ \\
\hline
$f_{M4}/\Lambda^{4}$  & [-7.38;7.51] $\times 10^{2}$ & [-4.13;4.25] $\times 10^{2}$ & [-3.91;3.62] $ \times 10^{2}$ & [-2.27;1.97] $ \times 10^{2}$ \\
\hline
$f_{M5}/\Lambda^{4}$  & [-9.07;8.73] $\times 10^{2}$ & [-5.17;4.84] $\times 10^{2}$ & [-6.98;6.73] $ \times 10^{2}$ & [-3.98;3.73] $ \times 10^{2}$ \\
\hline
$f_{M7}/\Lambda^{4}$  & [-3.27;3.19] $\times 10^{3}$ & [-1.85;1.77] $\times 10^{3}$ & [-2.46;2.49] $ \times 10^{3}$ & [-1.38;1.41] $ \times 10^{3}$ \\
\hline
$f_{T0}/\Lambda^{4}$  & [-1.26;1.20] $\times 10^{2}$ & [-0.73;0.66] $\times 10^{2}$ & [-7.46;7.37] $ \times 10^{1}$ & [-4.21;4.12] $ \times 10^{1}$ \\
\hline
$f_{T1}/\Lambda^{4}$  & [-7.70;7.78] $\times 10^{1}$ & [-4.31;4.39] $\times 10^{1}$ & [-6.19;5.90] $ \times 10^{1}$ & [-3.55;3.26] $ \times 10^{1}$ \\
\hline
$f_{T2}/\Lambda^{4}$  & [-1.92;2.32] $\times 10^{2}$ & [-1.01;1.41] $\times 10^{2}$ & [-1.34;1.86] $ \times 10^{2}$ & [-0.67;1.18] $ \times 10^{2}$ \\
\hline
$f_{T5}/\Lambda^{4}$  & [-3.65;3.92] $\times 10^{1}$ & [-2.00;2.27] $\times 10^{1}$ & [-2.10;2.36] $ \times 10^{1}$ & [-1.13;1.39] $ \times 10^{1}$ \\
\hline
$f_{T6}/\Lambda^{4}$  & [-2.15;2.61] $\times 10^{1}$ & [-1.12;1.59] $\times 10^{1}$ & [-1.67;2.03] $ \times 10^{1}$ & [-0.87;1.23] $ \times 10^{1}$ \\
\hline
$f_{T7}/\Lambda^{4}$  & [-0.71;0.58] $\times 10^{2}$ & [-0.43;0.30] $\times 10^{2}$ & [-4.87;4.73] $ \times 10^{1}$ & [-2.77;2.63] $ \times 10^{1}$ \\
\hline
\multicolumn{5}{|c|}{LHeC, \hspace{5mm} $\sqrt{s}$ = 1.98 TeV} \\
\hline
\cline{1-5}
$f_{M0}/\Lambda^{4}$  & [-3.84;3.51] $\times 10^{2}$ & [-2.24;1.91] $\times 10^{2}$  & [-2.40;2.45] $ \times 10^{2}$ & [-1.34;1.39] $\times 10^{2}$ \\
\hline
$f_{M1}/\Lambda^{4}$  & [-4.57;4.77] $\times 10^{2}$ & [-2.53;2.73]$ \times 10^{2}$  & [-4.28;4.30] $ \times 10^{2}$ & [-2.40;2.42] $\times 10^{2}$  \\
\hline
$f_{M2}/\Lambda^{4}$  & [-5.42;5.84] $\times 10^{1}$ & [-2.96;3.38]$ \times 10^{1}$  & [-3.67;3.80] $ \times 10^{1}$ & [-2.04;2.16] $\times 10^{1}$  \\
\hline
$f_{M3}/\Lambda^{4}$  & [-7.10;7.24] $\times 10^{1}$ & [-3.96;4.10]$ \times 10^{1}$  & [-6.42;6.69] $ \times 10^{1}$ & [-3.55;3.83] $\times 10^{1}$  \\
\hline
$f_{M4}/\Lambda^{4}$  & [-2.01;2.03] $\times 10^{2}$ & [-1.13;1.15]$ \times 10^{2}$  & [-1.37;1.33] $ \times 10^{2}$ & [-7.78;7.41] $\times 10^{1}$  \\
\hline
$f_{M5}/\Lambda^{4}$  & [-2.77;2.43] $\times 10^{2}$ & [-1.64;1.30]$ \times 10^{2}$  & [-2.51;2.28] $ \times 10^{2}$ & [-1.46;1.24] $\times 10^{2}$  \\
\hline
$f_{M7}/\Lambda^{4}$  & [-9.56;9.18] $\times 10^{2}$ & [-5.46;5.08]$ \times 10^{2}$  & [-8.48;8.67] $ \times 10^{2}$ & [-4.73;4.92] $\times 10^{2}$  \\
\hline
$f_{T0}/\Lambda^{4}$  & [-2.22;2.35] $\times 10^{1}$ & [-1.22;1.36] $\times 10^{1}$  & [-2.42;2.26] $ \times 10^{1}$ & [-1.40;1.24] $\times 10^{1}$ \\
\hline
$f_{T1}/\Lambda^{4}$  & [-2.02;2.18] $\times 10^{1}$ & [-1.10;1.26] $\times 10^{1}$  & [-1.99;1.94] $ \times 10^{1}$ & [-1.13;1.08] $\times 10^{1}$  \\
\hline
$f_{T2}/\Lambda^{4}$  & [-5.46;6.07] $\times 10^{1}$ & [-2.95;3.56] $\times 10^{1}$  & [-0.45;0.59] $ \times 10^{2}$ & [-0.23;0.37] $\times 10^{2}$ \\
\hline
$f_{T5}/\Lambda^{4}$  & [-9.64;10.09] & [-5.33;5.78]  & [-7.12;7.13]                  & [-4.00;4.01]   \\
\hline
$f_{T6}/\Lambda^{4}$  & [-6.22;6.64]                 & [-3.41;3.83]                  & [-6.07;5.83]                  & [-3.46;3.23]   \\
\hline
$f_{T7}/\Lambda^{4}$  & [-1.95;1.59] $\times 10^{1}$ & [-1.18;0.83] $\times 10^{1}$  & [-1.55;1.58] $ \times 10^{1}$ & [-8.67;8.94] \\
\hline \hline
\end{tabular}
\end{table}

\begin{table}
\caption{Sensitivity measures on the aQGC at the $95\%$ C. L. via the process $e^- p \to e^-\gamma^* p \to p W^- \gamma \nu_e$
at the FCC-he with $\sqrt{s}$ = 3.46, 5.29 TeV. Here, we consider that only one of the anomalous couplings deviates from the
SM at any given time.}
\begin{tabular}{|c|c|c|c|c|c|}
\hline \hline
\multicolumn{5}{|c|}{FCC-he, \hspace{5mm} $\sqrt{s}$ = 3.46 TeV}\\
\hline
& \multicolumn{2}{|c|}{Leptonic channel} & \multicolumn{2}{|c|}{ Hadronic channel }\\
\hline
Couplings (TeV$^{-4}$)&       100 fb$^{-1}$           & 1000 fb$^{-1}$                & 100 fb$^{-1}$                  & 1000 fb$^{-1}$ \\
\hline \cline{1-5}
$f_{M0}/\Lambda^{4}$  & [-2.98;2.97] $\times 10^{2}$ & [-1.68;1.67] $\times 10^{2}$ & [-4.69;4.88] $ \times 10^{1}$ & [-2.60;2.78] $ \times 10^{1}$ \\
\hline
$f_{M1}/\Lambda^{4}$  & [-3.56;3.60] $\times 10^{2}$ & [-2.00;2.03]$ \times 10^{2}$ & [-1.56;1.67] $ \times 10^{2}$ & [-0.86;0.97] $ \times 10^{2}$ \\
\hline
$f_{M2}/\Lambda^{4}$  & [-4.54;4.58] $\times 10^{1}$ & [-2.54;2.58]$ \times 10^{1}$ & [-7.10;7.55]                  & [-3.90;4.35]  \\
\hline
$f_{M3}/\Lambda^{4}$  & [-5.35;5.57] $\times 10^{1}$ & [-2.96;3.18]$ \times 10^{1}$ & [-2.57;2.40] $ \times 10^{1}$ & [-1.48;1.31] $ \times 10^{1}$ \\
\hline
$f_{M4}/\Lambda^{4}$  & [-1.60;1.69] $\times 10^{2}$ & [-0.88;0.97]$ \times 10^{2}$ & [-2.69;2.61] $ \times 10^{1}$ & [-1.53;1.45] $ \times 10^{1}$ \\
\hline
$f_{M5}/\Lambda^{4}$  & [-2.05;1.91] $\times 10^{2}$ & [-1.18;1.04]$ \times 10^{2}$ & [-9.03;8.86] $ \times 10^{1}$ & [-5.11;4.95] $ \times 10^{1}$ \\
\hline
$f_{M7}/\Lambda^{4}$  & [-7.13;7.12] $\times 10^{2}$ & [-4.01;4.00]$ \times 10^{2}$ & [-3.03;3.48] $ \times 10^{2}$ & [-1.61;2.07] $ \times 10^{2}$ \\
\hline
$f_{T0}/\Lambda^{4}$  & [-1.66;1.65] $\times 10^{1}$ & [-9.34;9.23] & [-3.40;4.03]                  & [-1.79;2.42]  \\
\hline
$f_{T1}/\Lambda^{4}$  & [-1.05;1.34] $\times 10^{1}$ & [-0.54;0.83] $\times 10^{1}$                & [-0.61;0.48] $ \times 10^{1}$ & [-0.38;0.25] $\times 10^{1}$ \\
\hline
$f_{T2}/\Lambda^{4}$  & [-0.27;0.38] $\times 10^{2}$ & [-0.13;0.24] $\times 10^{2}$ & [-0.91;1.27] $ \times 10^{1}$ & [-0.45;0.81] $ \times 10^{1}$ \\
\hline
$f_{T5}/\Lambda^{4}$  & [-5.02;5.09]                 & [-2.81;2.88]                 & [-1.01;1.26]                  & [-0.52;0.77]  \\
\hline
$f_{T6}/\Lambda^{4}$  & [-3.57;3.70]                 & [-1.98;2.11]                 & [-1.37;2.01]                  & [-0.67;1.31]  \\
\hline
$f_{T7}/\Lambda^{4}$  & [-1.05;0.92] $\times 10^{1}$ & [-0.62;0.49] $\times 10^{1}$ & [-3.36;3.21]                  & [-1.92;1.78]  \\
\hline
\multicolumn{5}{|c|}{FCC-he, \hspace{5mm}   $\sqrt{s}$ = 5.29 TeV}\\
\hline\cline{1-5}
$f_{M0}/\Lambda^{4}$  & [-5.27;5.14] $\times 10^{1}$ & [-2.99;2.86] $\times 10^{1}$ & [-2.08;2.13] $ \times 10^{1}$ & [-1.16;1.21] $\times 10^{1}$ \\
\hline
$f_{M1}/\Lambda^{4}$  & [-5.86;5.94] $\times 10^{1}$ & [-3.28;3.36]$ \times 10^{1}$ & [-6.91;7.09] $ \times 10^{1}$ & [-3.86;4.02] $\times 10^{1}$ \\
\hline
$f_{M2}/\Lambda^{4}$  & [-7.92;8.10]                 & [-4.42;4.60]                 & [-3.15;3.25]                  & [-1.75;1.85]  \\
\hline
$f_{M3}/\Lambda^{4}$  & [-8.67;8.99] & [-4.81;5.13]                 & [-1.01;1.13] $ \times 10^{1}$ & [-0.55;0.66] $\times 10^{1}$ \\
\hline
$f_{M4}/\Lambda^{4}$  & [-2.75;2.92] $\times 10^{1}$ & [-1.51;1.68]$ \times 10^{1}$ & [-1.20;1.12] $ \times 10^{1}$ & [-6.96;6.12] \\
\hline
$f_{M5}/\Lambda^{4}$  & [-3.32;3.19] $\times 10^{1}$ & [-1.89;1.77]$ \times 10^{1}$ & [-3.95;3.84] $ \times 10^{1}$ & [-2.24;2.14] $\times 10^{1}$ \\
\hline
$f_{M7}/\Lambda^{4}$  & [-1.23;1.11] $\times 10^{2}$ & [-0.72;0.60]$ \times 10^{2}$ & [-1.40;1.42] $ \times 10^{2}$ & [-7.83;8.01] $\times 10^{1}$  \\
\hline
$f_{T0}/\Lambda^{4}$  & [-2.90;2.88]                 & [-1.64;1.62]                 & [-1.51;1.43]                  & [-0.86;0.79] \\
\hline
$f_{T1}/\Lambda^{4}$  & [-1.69;1.82]                 & [-0.92;1.05]                 & [-2.21;2.10]                  & [-1.27;1.16] \\
\hline
$f_{T2}/\Lambda^{4}$  & [-4.56;5.00]                 & [-2.47;2.92]                 & [-3.92;4.76]                  & [-2.04;2.88] \\
\hline
$f_{T5}/\Lambda^{4}$  & [-0.83;0.96]                 & [-0.44;0.57]                 & [-4.33;4.66] $ \times 10^{-1}$& [-2.37;2.70] $\times 10^{-1}$ \\
\hline
$f_{T6}/\Lambda^{4}$  & [-0.49;0.58]& [-0.26;0.35] & [-0.61;0.70]                  & [-0.33;0.42] \\
\hline
$f_{T7}/\Lambda^{4}$  & [-1.63;1.35]                 & [-0.98;0.71]                 & [-1.57;1.12]                  & [-1.00;0.55] \\
\hline
\end{tabular}
\end{table}

\begin{table}
\caption{Sensitivity measures on the aQGC at the $95\%$ C. L. via the process $e^- p \to e^-\gamma^* p \to p W^- \gamma \nu_e$
at the FCC-he with $\sqrt{s}$ = 3.46, 5.29 TeV and $1000fb^{-1}$ for the combined data. Here, we consider that only one of the
anomalous couplings deviates from the
SM at any given time.}
\begin{tabular}{|c|c|c|}\hline
Couplings   & $\sqrt{s} = 3.46 $ TeV & $\sqrt{s} = 5.29 $ TeV \\
\hline \hline
$f_{M0}/\Lambda^{4}$  & [-2.25;2.43] $ \times 10^{1}$ & [-1.00;1.03] $ \times 10^{1}$\\
\hline
$f_{M1}/\Lambda^{4}$  &[-0.71;0.82] $ \times 10^{2}$ & [-2.61;2.75] $ \times 10^{1}$ \\
\hline
$f_{M2}/\Lambda^{4}$  & [-3.36;3.81] & [-1.49;1.59] \\
\hline
$f_{M3}/\Lambda^{4}$  & [-1.24;1.10] $ \times 10^{1}$ & [-3.68;4.46] \\
\hline
$f_{M4}/\Lambda^{4}$  & [-1.33;1.26] $ \times 10^{1}$ & [-5.91;5.27] \\
\hline
$f_{M5}/\Lambda^{4}$  & [-4.36;4.09] $ \times 10^{1}$ & [-1.54;1.43] $ \times 10^{1}$ \\
\hline
$f_{M7}/\Lambda^{4}$  & [-1.34;1.76] $ \times 10^{2}$  & [-5.56;5.15] $ \times 10^{1}$ \\
\hline
$f_{T0}/\Lambda^{4}$  & [-2.13;1.52] & [-7.26;6.59] $ \times 10^{-1}$  \\
\hline
$f_{T1}/\Lambda^{4}$  & [-3.12;2.09] $ \times 10^{1}$ & [-8.15;8.10] $ \times 10^{-1}$  \\
\hline
$f_{T2}/\Lambda^{4}$  & [-0.35;0.75] $ \times 10^{1}$ & [-1.49;2.22]  \\
\hline
$f_{T5}/\Lambda^{4}$  & [-0.43;0.69]  & [-1.91;2.35] $ \times 10^{-1}$   \\
\hline
$f_{T6}/\Lambda^{4}$  & [-0.54;1.14] & [-0.21;0.29] \\
\hline
$f_{T7}/\Lambda^{4}$  & [-1.68;1.48]  & [-0.79;0.39]  \\
\hline
\end{tabular}
\end{table}

Photon in the final state of the process $e^- p \to e^-\gamma^* p \to p W^- \gamma \nu_e$ at the LHeC and FCC-he have the advantage
of being identifiable with high purity and efficiency. Thus, the single-photon and diphoton channels are especially sensitive for new
physics BSM in terms of modest backgrounds, excellent mass resolution and the clean experimental signature. Furthermore,
as we mentioned above, in order to quantify the expected limits on $f_{M,i}/\Lambda^4$ and $f_{T,j}/\Lambda^4$, advantage has been
taken in this analysis of the fact that the aQGC enhance the cross-sections at high energies. In Figs. 11 and 12, the number of expected events as a function of the $p_T(\gamma)$ transverse momentum for the $e^- p \to e^-\gamma^* p \to p W^- \gamma \nu_e$ signal and backgrounds at the LHeC and the FCC-he. The distributions are for $\frac{f_{M,i}}{\Lambda^4}$ with $i=0,1,2, 3,4,5,7$ , $\frac{f_{T,i}}{\Lambda^4}$ with $j=0,1,2,5,6,7$ and various backgrounds for both leptonic and hadronic decay channel of the W-boson. 
In these figures, the solid histograms correspond to $\sigma_{Tot}$ cross-section, and the dashed distribution corresponds to the other SM background sources. In addition, this distribution clearly shows great sensitivity with respect to the anomalous $f_{M,i}/\Lambda^4$ and $f_{T,j}/\Lambda^4$ couplings for both cases leptonic
and hadronic. The analysis of these distributions is important to be able to discriminate the basic acceptance cuts for $W^-\gamma\nu_e$
events at the LHeC and the FCC-he.

\section{Summary and conclusions}

In this work, in the effective Lagrange approach, we study the $e^- p \to e^-\gamma^* p \to p W^- \gamma \nu_e$ channel
at the LHeC and the FCC-he as a way to perform sensitivity measures on the total cross-section and on the anomalous
$f_{M,i}/\Lambda^4$ and $f_{T,j}/\Lambda^4$ couplings defining dimension-8 effective operators. Since the aQGC $WW\gamma\gamma$
described through effective Lagrangian have dimension-8, they have very strong energy dependence.
Therefore, the anomalous cross-section containing the $WW\gamma\gamma$ vertex has a higher energy than the SM cross-section.
In addition, the future $ep$ collider will possibly generate a final state with two or more gauge bosons.
Hence, it will have a great potential to investigate the aQGC. High energy accelerated $e^-$ and $p$ beams at these colliders radiate
quasi-real photons, and thus $e\gamma^*$, $\gamma^* p$ and $\gamma^*\gamma^*$ collisions are produced from the $e^-p$ process itself.
Therefore, $ep$ colliders will provide an important opportunity to probe $e\gamma^*$, $\gamma^* p$ and $\gamma^*\gamma^*$ collisions
at high energies. For the new physics searches at $ep$ colliders have a very clean experimental environment, since
they have no interference with weak and strong interactions.

Regarding the comparison with present experimental limits, we find that the $f_{M,i}/\Lambda^4$ and $f_{T,j}/\Lambda^4$
constraints are significantly competitive with the ones achievable at the CMS and ATLAS Collaborations at the LHC through
the $Z\gamma jj$ production \cite{CMS-2002.09902}. Specifically, our results provide stringent limits on the parameters
$f_{T,5}/\Lambda^4$, $f_{T,6}/\Lambda^4$ and $f_{T,7}/\Lambda^4$ by a factor of ${\cal O}(1.9-3.7)$ than those reported
in Ref. \cite{CMS-2002.09902}. Our limits stronger at $95\%$ C.L. on the Wilson coefficients are
$\frac{f_{T5}}{\Lambda^{4}}= [-0.237; 0.270] \hspace{1mm} {\rm TeV^{-4}}$, $\frac{f_{T6}}{\Lambda^{4}}= [-0.330; 0.420]
\hspace{1mm} {\rm TeV^{-4}}$ and $\frac{f_{T7}}{\Lambda^{4}}= [-1.000; 0.550] \hspace{1mm} {\rm TeV^{-4}}$ for $\sqrt{s}=5.29$ TeV
and ${\cal L}=1000$ ${\rm fb^{-1}}$ at $95\%$ C.L. for the FCC-he.

We conclude by mentioning that our projections at the LHeC and the FCC-he are interpreted in the approach of dimension-8
effective field theory operators through the $e^- p \to e^-\gamma^* p \to p W^- \gamma \nu_e$ channel. Our results indicate
that the $e^- p \to e^-\gamma^* p \to p W^- \gamma \nu_e$ production is convincing for searching for the dimension-8 operators
${\cal O}_{M,0-5,7}$ and ${\cal O}_{T,0-2,5,6,7}$, and as a consequence of the Wilson coefficients $f_{M,0-5,7}/\Lambda^4$
and $f_{T,0-2,5,6,7}/\Lambda^4$ with clean environments, as well as with good sensitivity.

\vspace{1cm}

\begin{center}
{\bf Acknowledgments}
\end{center}

A. G. R. and M. A. H. R. thank SNI and PROFEXCE (M\'exico). The numerical calculations reported in this paper were fully performed 
at TUBITAK ULAKBIM, High Performance and Grid Computing Center (TRUBA resources).


\newpage

\begin{figure}[t]
\centerline{\scalebox{0.7}{\includegraphics{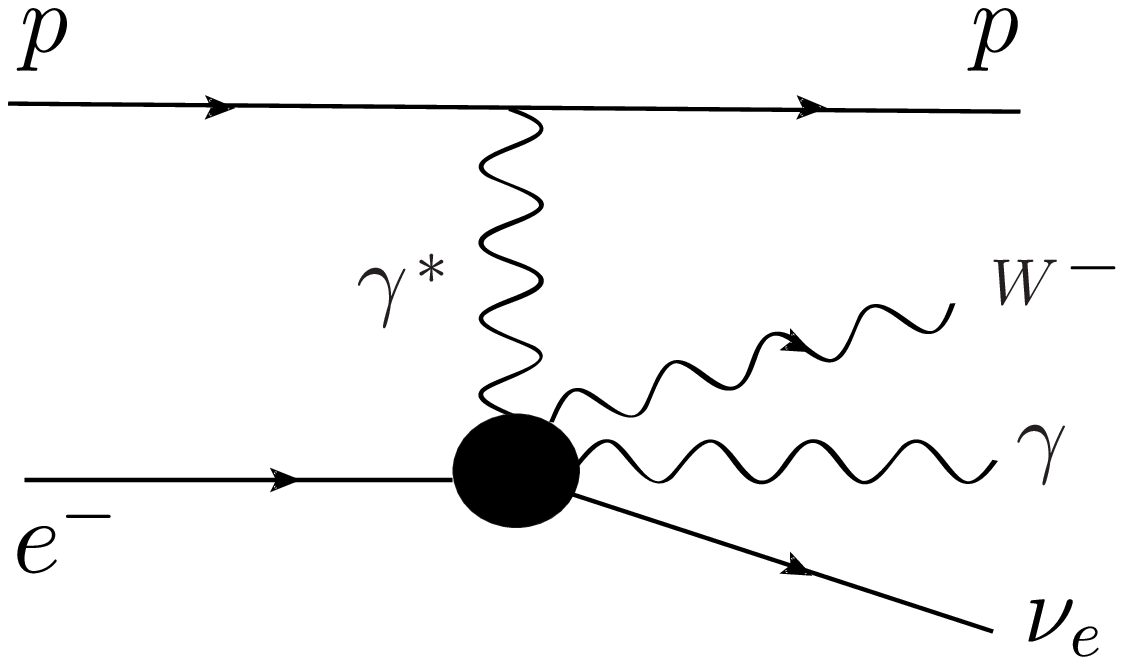}}}
\caption{ \label{fig:gamma1} Feynman diagram for the signal process $e^-p \to e^-\gamma^* p \to p W^-\gamma \nu_e $.
New physics (represented by a black circle) in the electroweak sector can modify the quartic gauge couplings.}
\label{Fig.1}
\end{figure}

\begin{figure}[t]
\centerline{\scalebox{0.8}{\includegraphics{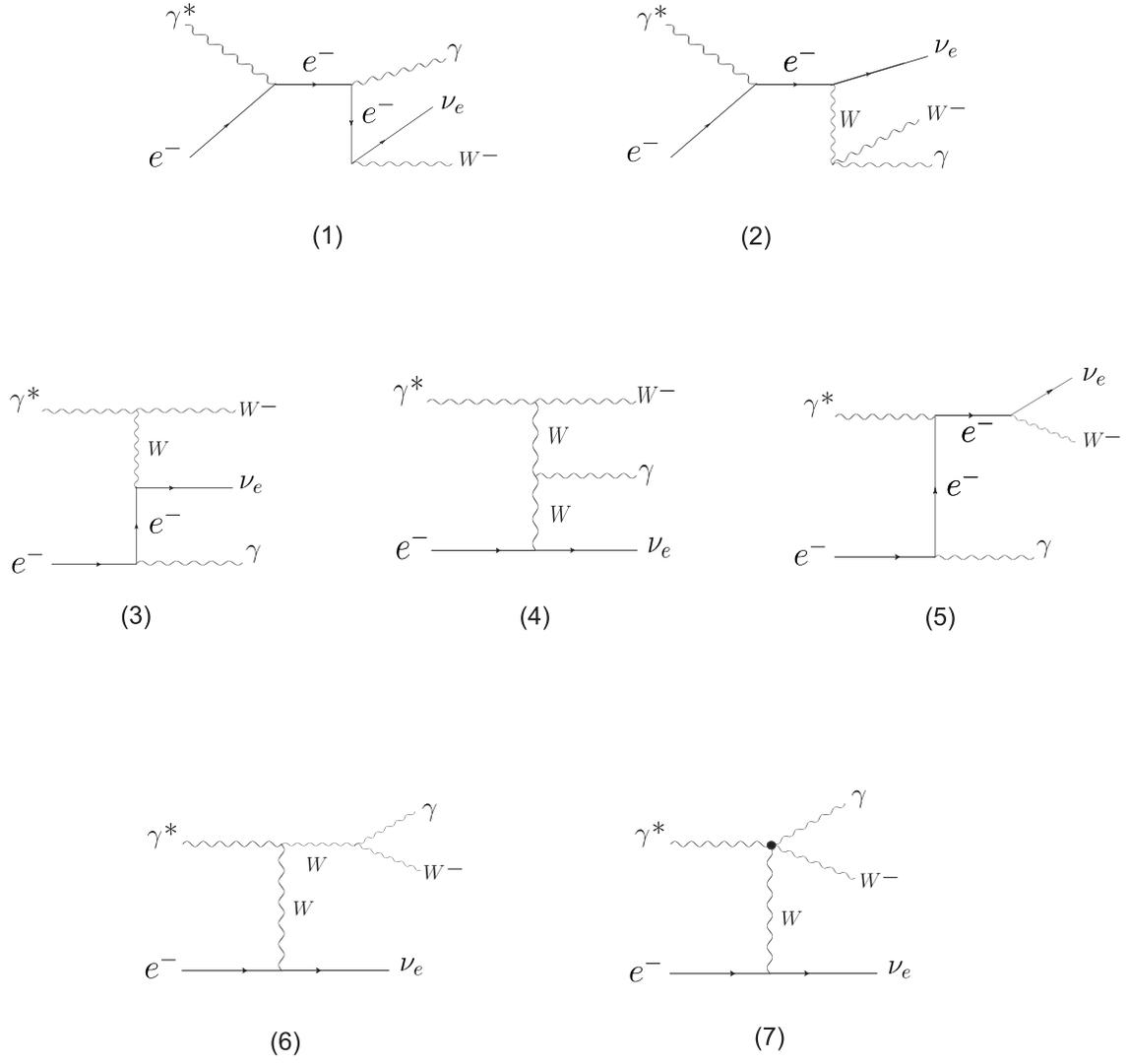}}}
\caption{ \label{fig:gamma2} Representative Feynman diagrams contributing to the subprocess
$e^-\gamma^* \to W^-\gamma \nu_e$.}
\label{Fig.2}
\end{figure}

\begin{figure}[t]
\centerline{\scalebox{1.2}{\includegraphics{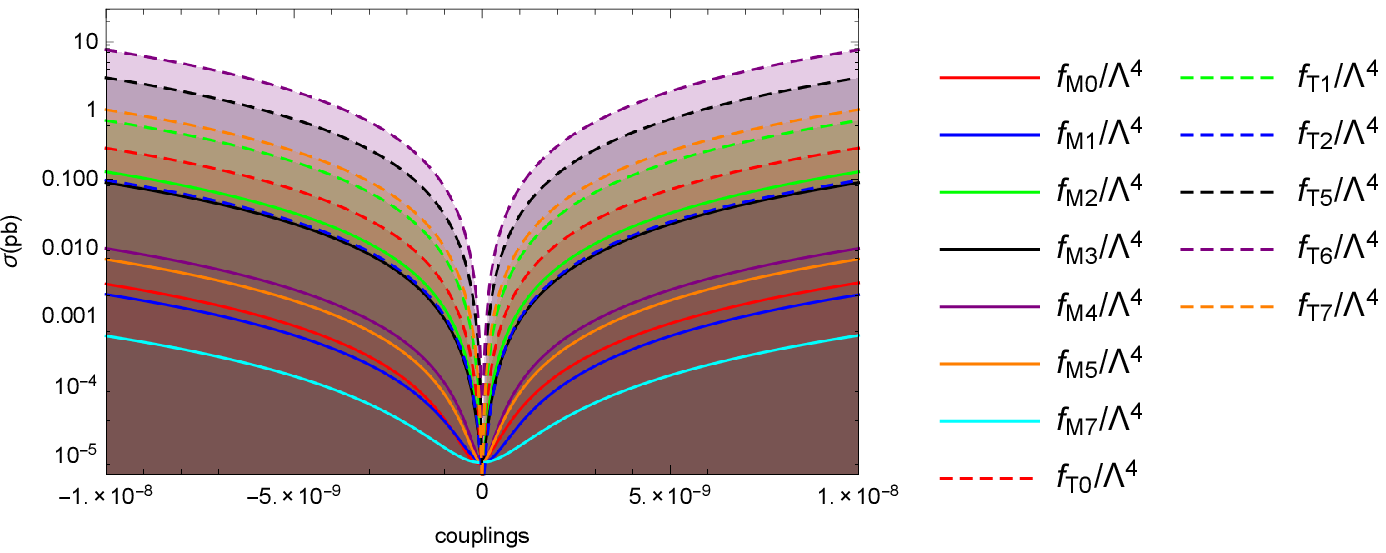}}}
\caption{ \label{fig:gamma1} For leptonic channel, the total cross-sections of the process
$e^-p \to e^-\gamma^* p \to p W^-\gamma \nu_e $ as a function of the anomalous couplings
at the LHeC with $\sqrt{s}=1.30$ TeV.}
\label{Fig.1}
\end{figure}

\begin{figure}[t]
\centerline{\scalebox{1.2}{\includegraphics{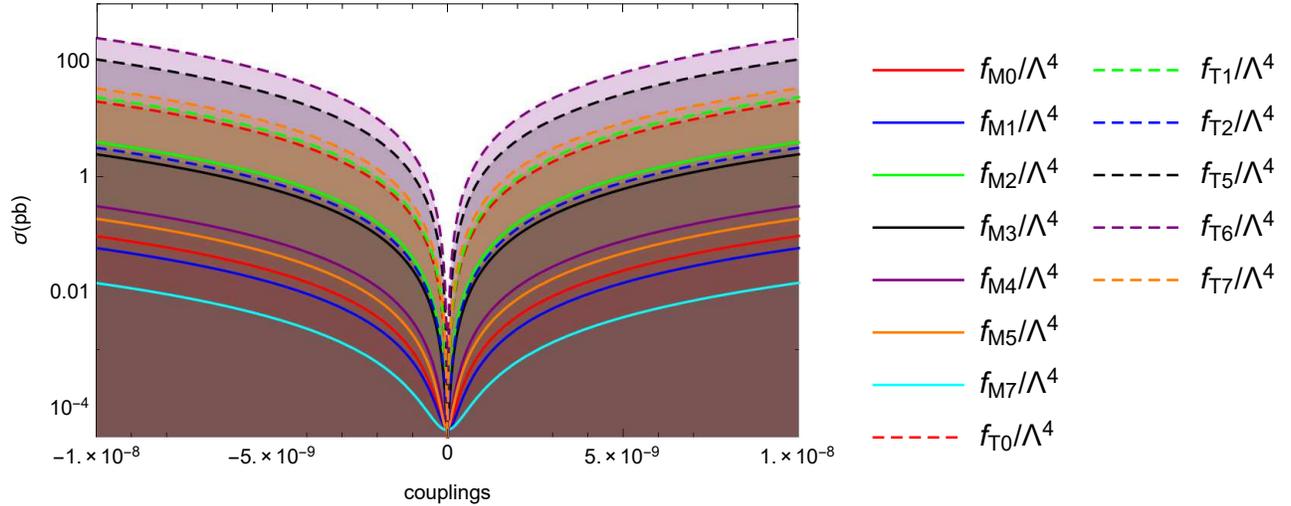}}}
\caption{ \label{fig:gamma2} Same as in Fig. 3, but for $\sqrt{s}=1.98$ TeV at the LHeC.}
\label{Fig.2}
\end{figure}

\begin{figure}[t]
\centerline{\scalebox{1.2}{\includegraphics{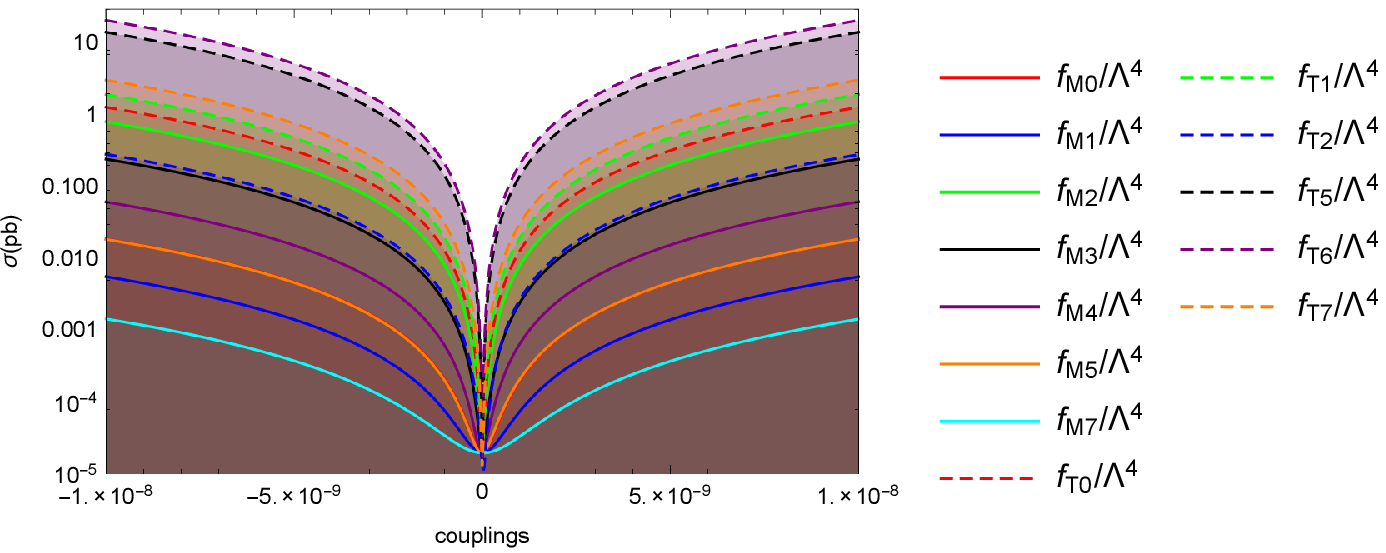}}}
\caption{ Same as in Fig. 3, but for hadronic decay.}
\label{Fig.3}
\end{figure}

\begin{figure}[t]
\centerline{\scalebox{1.2}{\includegraphics{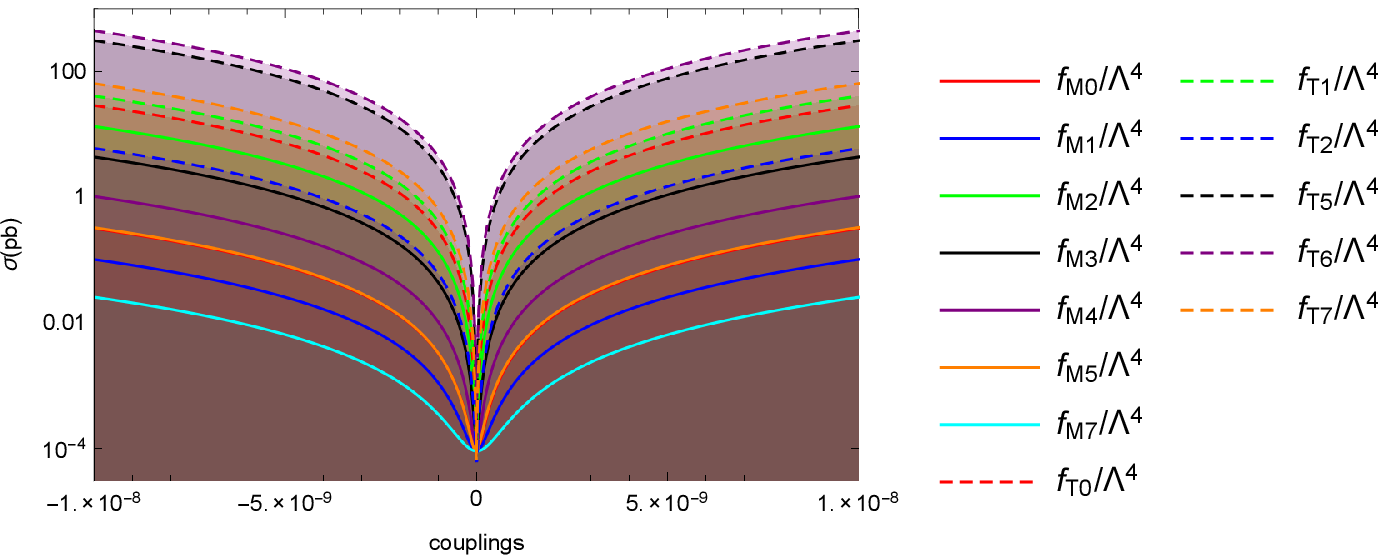}}}
\caption{ Same as in Fig. 4, but for hadronic decay.}
\label{Fig.4}
\end{figure}

\begin{figure}[t]
\centerline{\scalebox{1.2}{\includegraphics{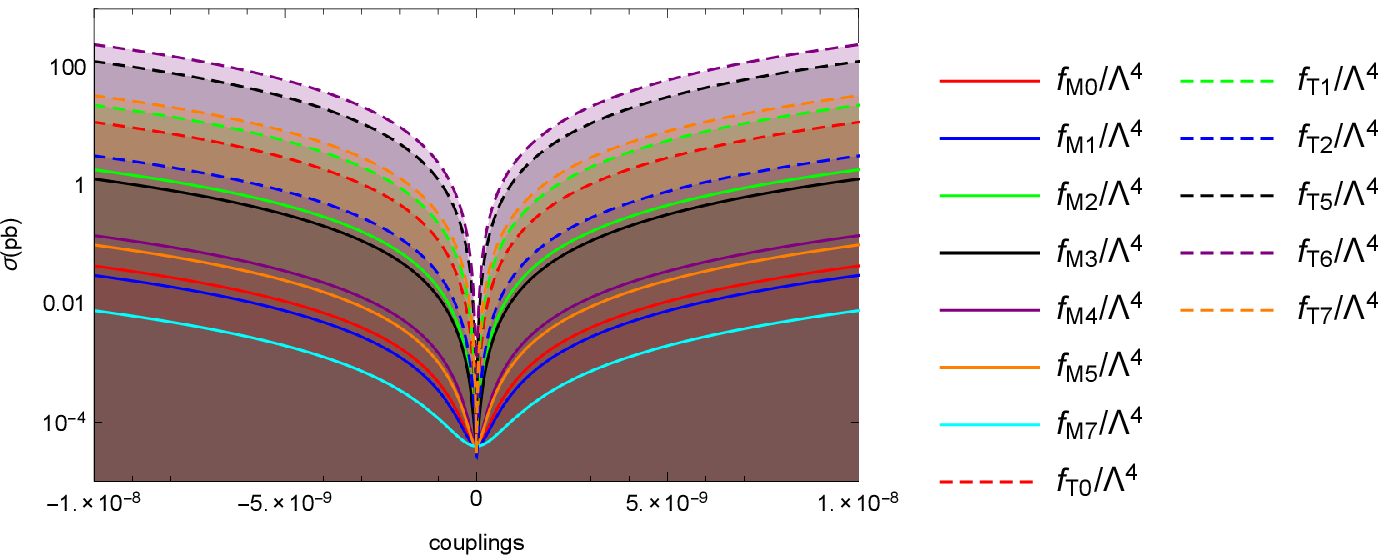}}}
\caption{ \label{fig:gamma1} For leptonic channel, the total cross-sections of the process
$e^-p \to e^-\gamma^* p \to p W^-\gamma \nu_e $ as a function of the anomalous couplings
at the FCC-he with $\sqrt{s}=3.46$ TeV.}
\label{Fig.1}
\end{figure}

\begin{figure}[t]
\centerline{\scalebox{1.2}{\includegraphics{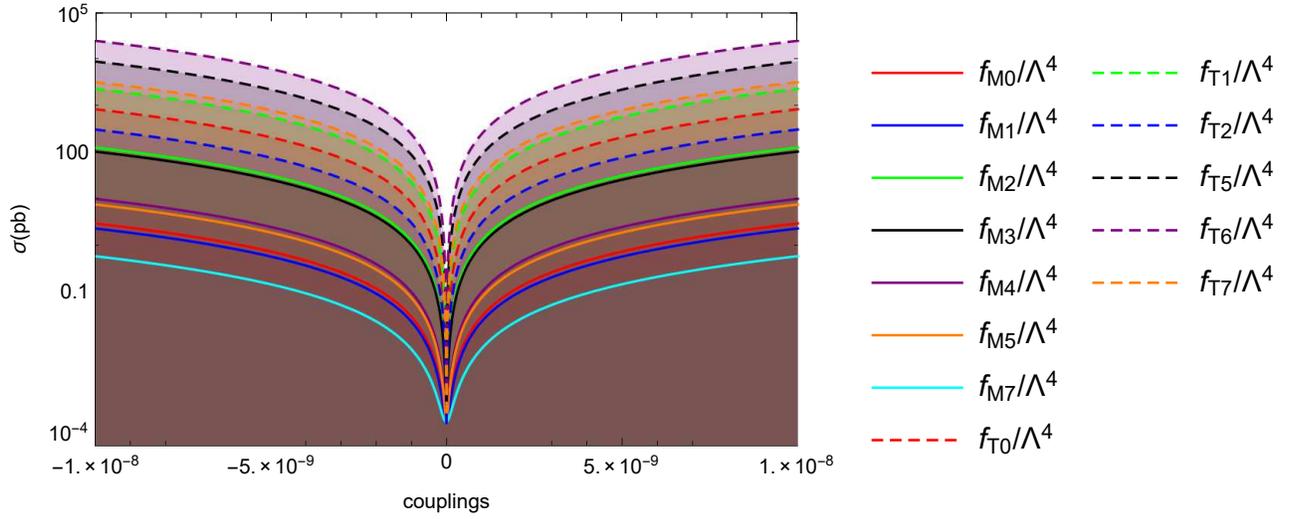}}}
\caption{ \label{fig:gamma2} Same as in Fig. 7, but for $\sqrt{s}=5.29$ TeV at the FCC-he.}
\label{Fig.2}
\end{figure}

\begin{figure}[t]
\centerline{\scalebox{1.2}{\includegraphics{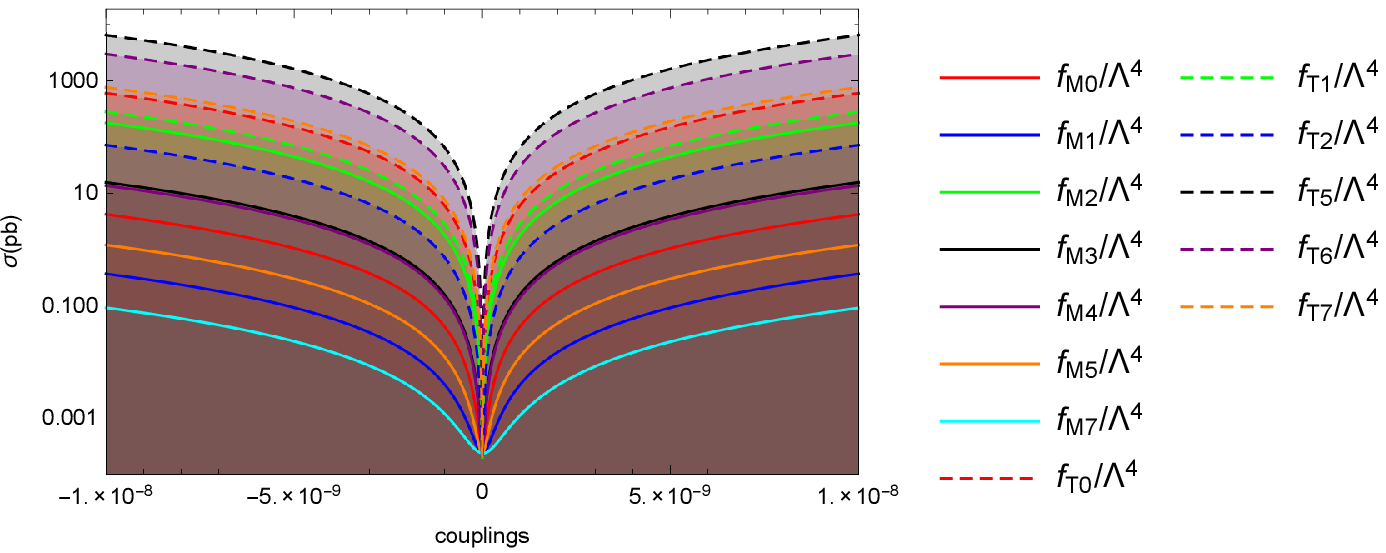}}}
\caption{ Same as in Fig. 7, but for hadronic decay.}
\label{Fig.3}
\end{figure}

\begin{figure}[t]
\centerline{\scalebox{1.2}{\includegraphics{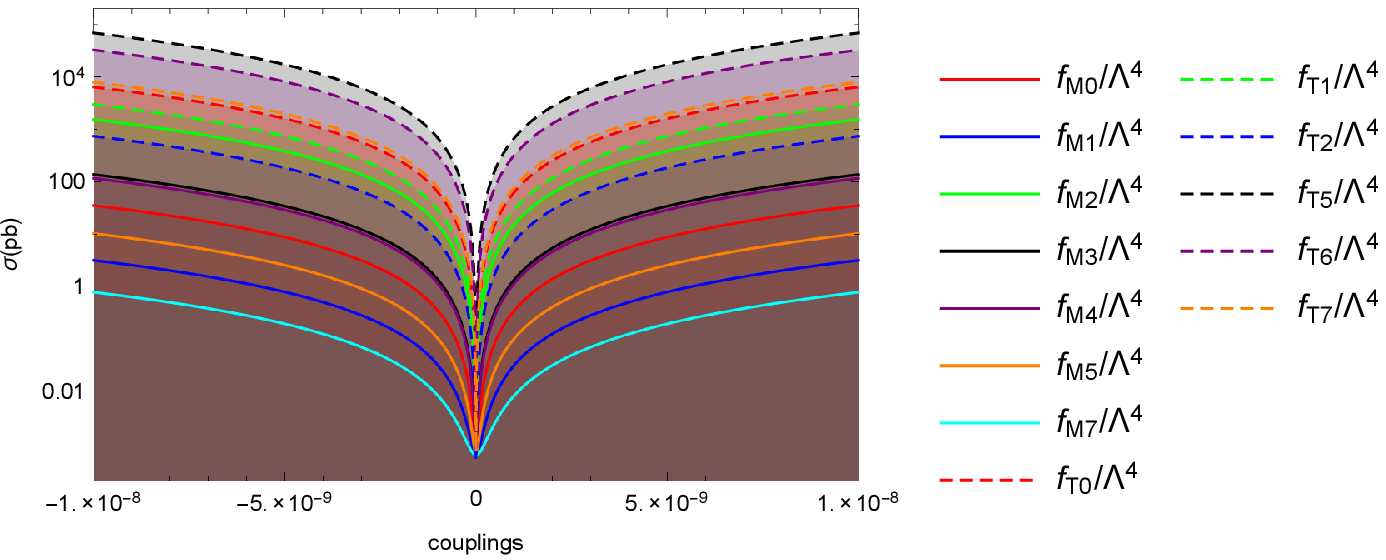}}}
\caption{ Same as in Fig. 8, but for hadronic decay.}
\label{Fig.4}
\end{figure}

\begin{figure}[ht]
  \begin{subfigure}[b]{0.5\linewidth}
    \centering
    \includegraphics[width=1.00\linewidth]{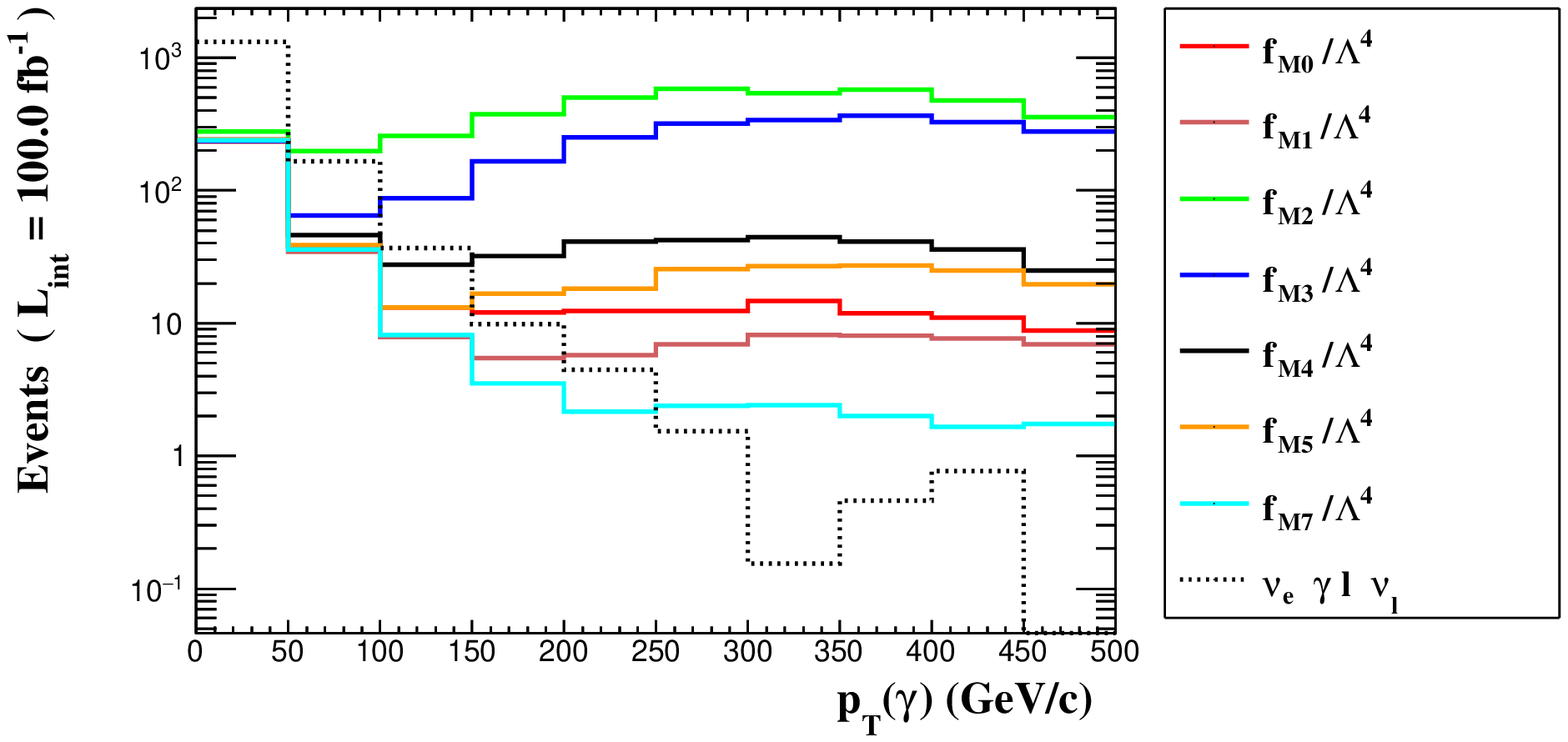}
    \caption{Leptonic decay}
    \label{fig7:a}
    \vspace{4ex}
  \end{subfigure}
  \begin{subfigure}[b]{0.5\linewidth}
    \centering
    \includegraphics[width=1.00\linewidth]{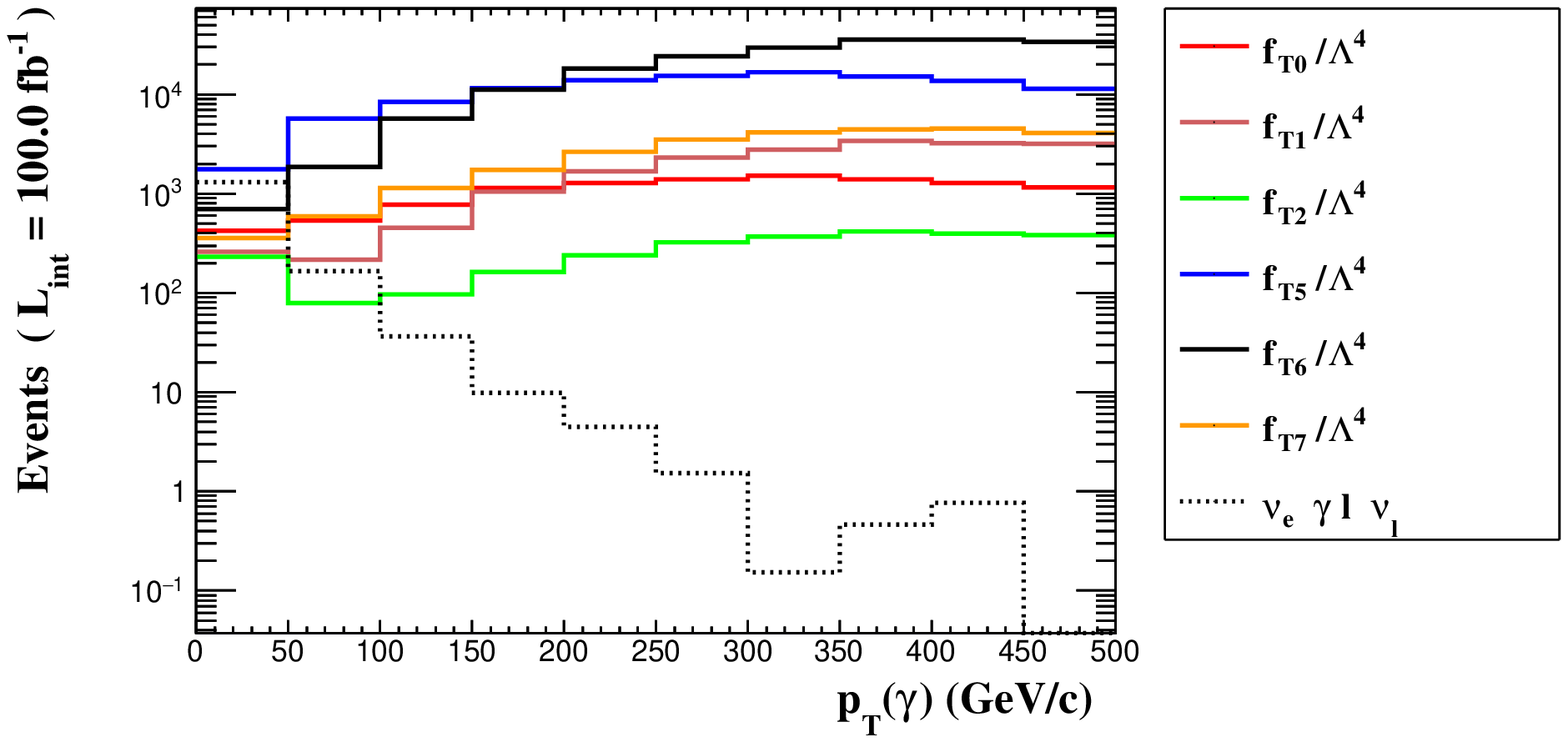}
    \caption{Leptonic decay}
    \label{fig7:b}
    \vspace{4ex}
  \end{subfigure}
  \begin{subfigure}[b]{0.5\linewidth}
    \centering
    \includegraphics[width=1.00\linewidth]{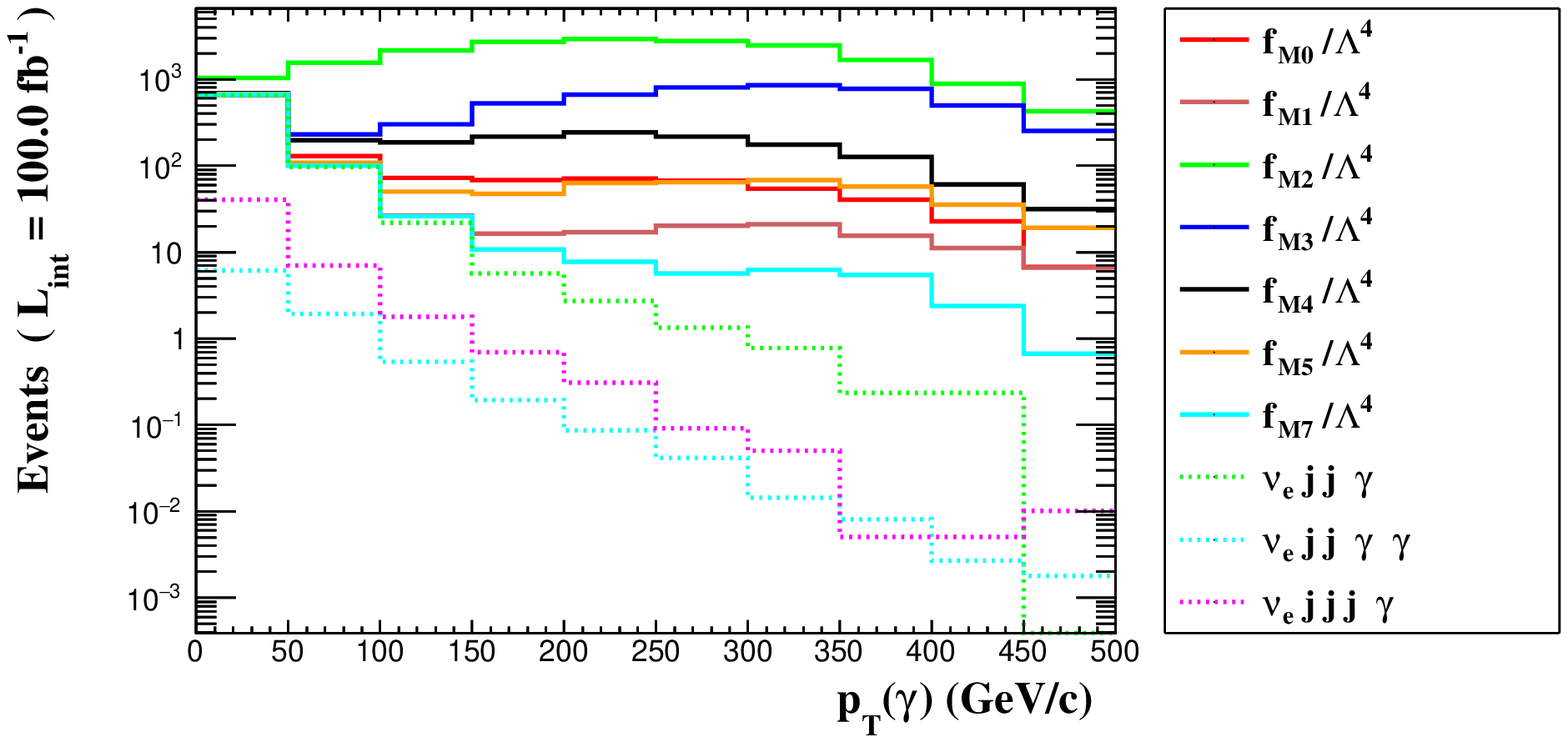}
    \caption{Hadronic decay}
    \label{fig7:c}
  \end{subfigure}
  \begin{subfigure}[b]{0.5\linewidth}
    \centering
    \includegraphics[width=1.00\linewidth]{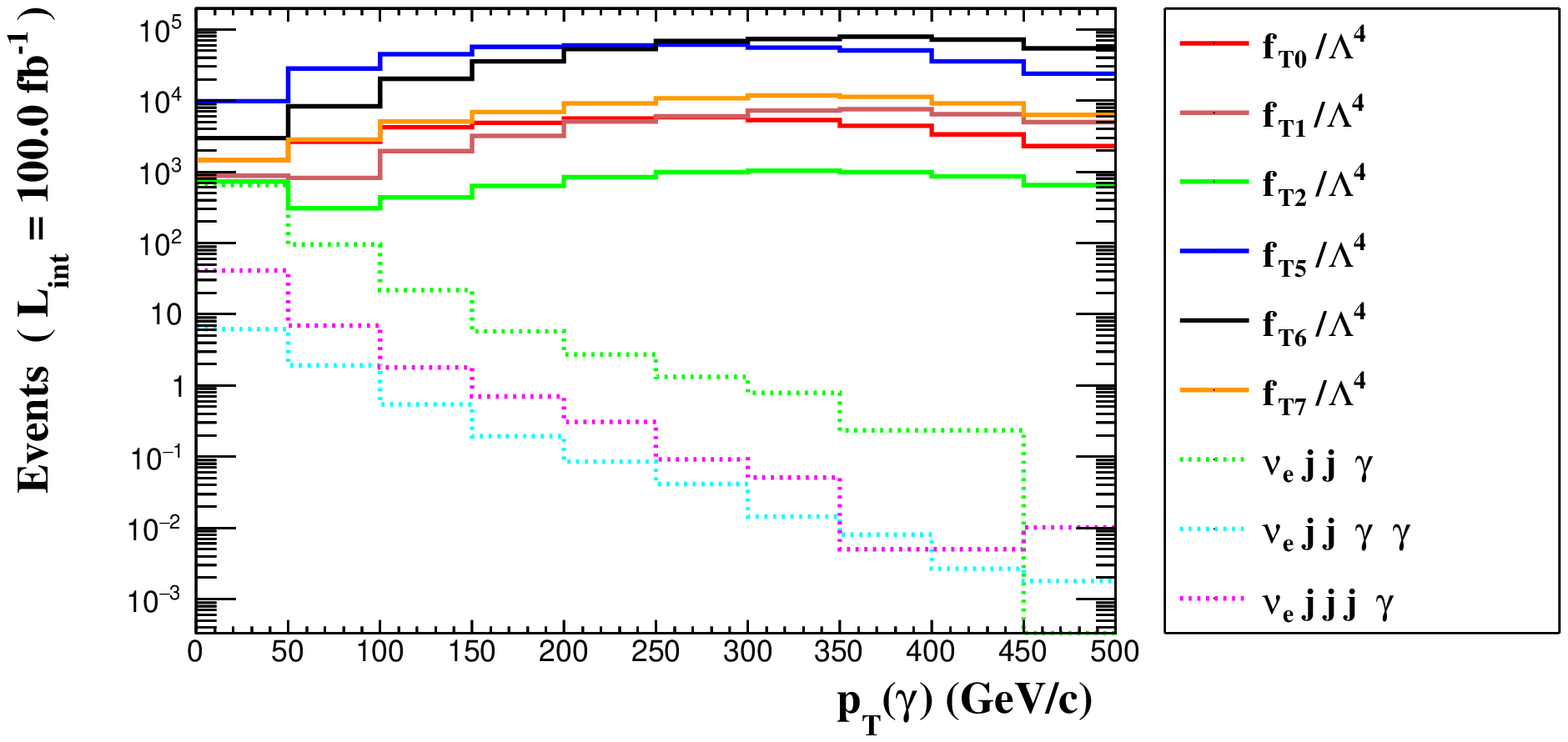}
    \caption{Hadronic decay}
    \label{fig7:d}
  \end{subfigure}
  \caption{The number of expected events as a function of the $p^{\gamma}_T$
photon transverse momentum for the $e^-p \to e^-\gamma^* p \to p W^-\gamma \nu_e$ signal and
backgrounds at the LHeC with $\sqrt{s}=1.98$ TeV. The distributions are for $f_{M,i}/\Lambda^4$
with $i=0,1,2,3,4,5,7$ , $f_{T,j}/\Lambda^4$  with $j=0,1,2,5,6,7$ and various backgrounds
for both leptonic and hadronic decay channel of the $W$-boson.}
  \label{Fig.3}
\end{figure}

\begin{figure}[ht]
  \begin{subfigure}[b]{0.5\linewidth}
    \centering
    \includegraphics[width=1.00\linewidth]{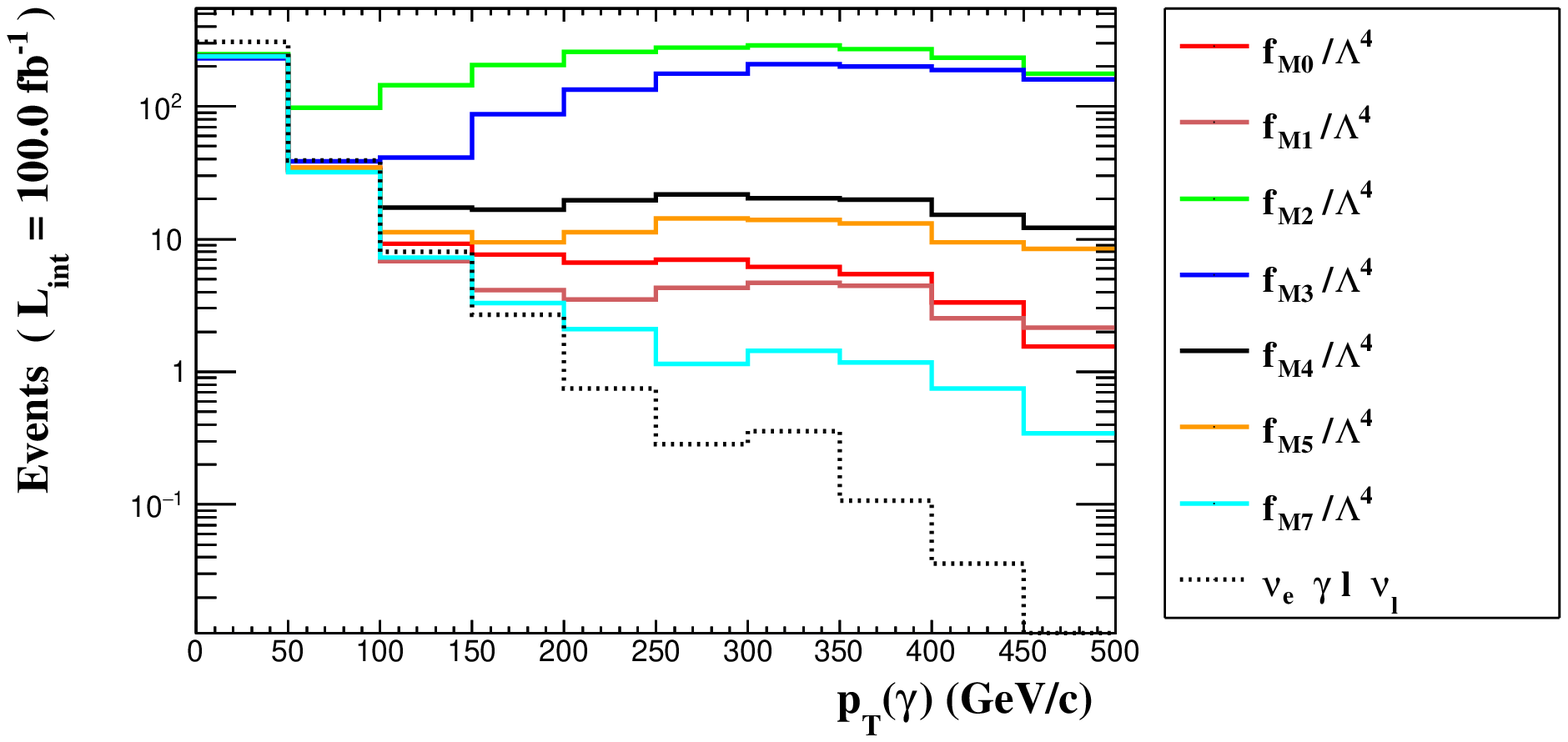}
    \caption{Leptonic decay}
    \label{fig8:a}
    \vspace{4ex}
  \end{subfigure}
  \begin{subfigure}[b]{0.5\linewidth}
    \centering
    \includegraphics[width=1.00\linewidth]{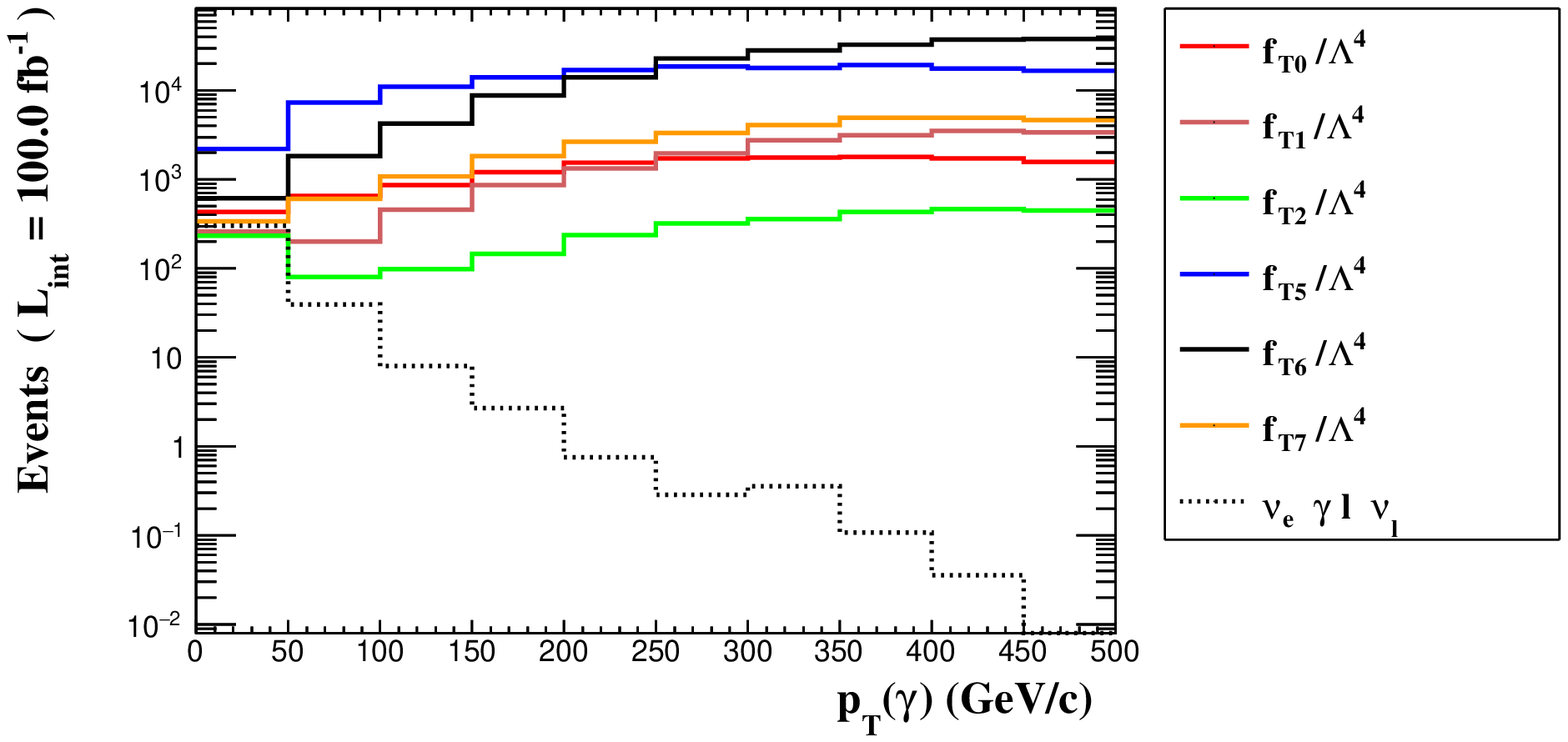}
    \caption{Leptonic decay}
    \label{fig8:b}
    \vspace{4ex}
  \end{subfigure}
  \begin{subfigure}[b]{0.5\linewidth}
    \centering
    \includegraphics[width=1.00\linewidth]{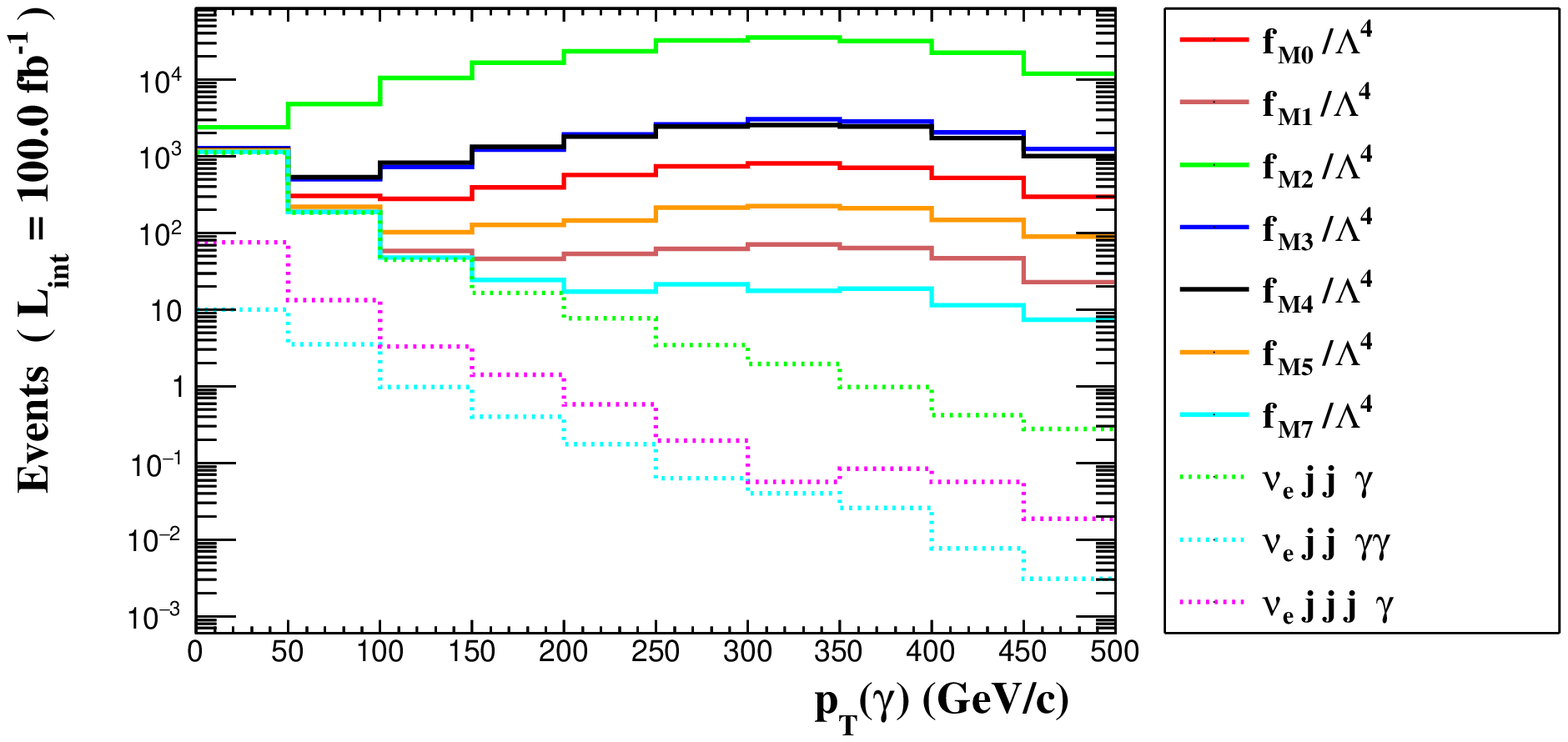}
    \caption{Hadronic decay}
    \label{fig8:c}
  \end{subfigure}
  \begin{subfigure}[b]{0.5\linewidth}
    \centering
    \includegraphics[width=1.00\linewidth]{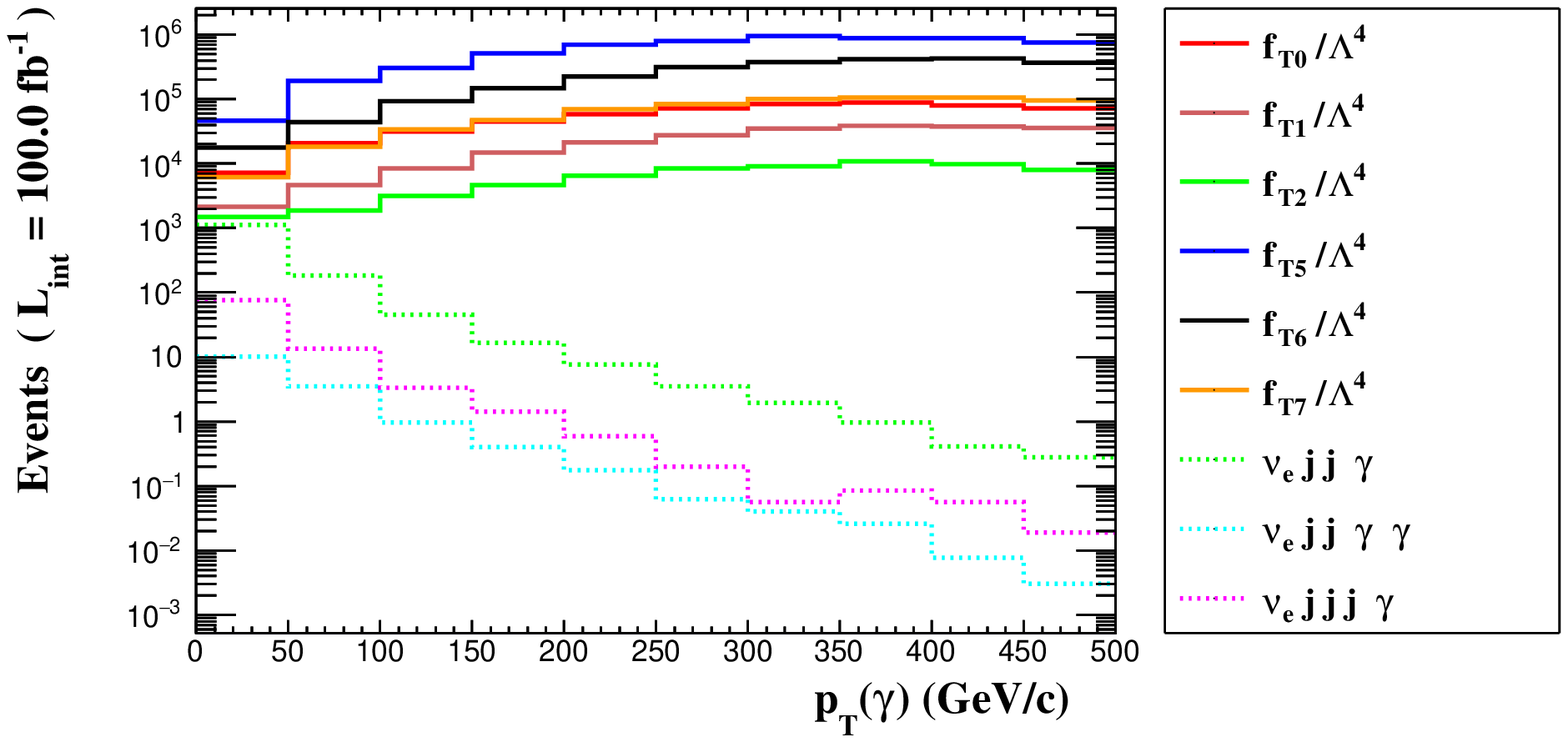}
    \caption{Hadronic decay}
    \label{fig8:d}
  \end{subfigure}
  \caption{The number of expected events as a function of the $p^{\gamma}_T$
photon transverse momentum for the $e^-p \to e^-\gamma^* p \to p W^-\gamma \nu_e$ signal and
backgrounds at the FCC-he with $\sqrt{s}=3.46$ TeV. The distributions are for $f_{M,i}/\Lambda^4$
with $i=0,1,2,3,4,5,7$ , $f_{T,j}/\Lambda^4$  with $j=0,1,2,5,6,7$ and various backgrounds
for both leptonic and hadronic decay channel of the $W$-boson.}
  \label{Fig.4}
\end{figure}

\end{document}